\documentclass[twocolumn]{aastex631}

\usepackage[T1]{fontenc}
\usepackage{romanbar}
\usepackage{appendix}
\usepackage{booktabs} 
\usepackage{float}
\usepackage{graphicx}
\usepackage{color}
 \usepackage{amsmath}
\usepackage{multirow}

\shortauthors{Huseby et al.}

\begin{document}

\title{Effects of Ultraviolet Radiation on Sub-Neptune Exoplanet Hazes Through Laboratory Experiments}


\author[0009-0009-5477-9375]{Lori Huseby}
\affiliation{Lunar and Planetary Laboratory, University of Arizona, Tucson, AZ 85721, USA}

\author[0000-0002-6721-3284]{Sarah E. Moran}
\affiliation{Lunar and Planetary Laboratory, University of Arizona, Tucson, AZ 85721, USA}
\affiliation{NASA Goddard Space Flight Center, Greenbelt, MD 20771, USA}
\affiliation{NHFP Sagan Fellow}

\author[0000-0002-0183-1581]{Neil Pearson}
\affiliation{Lunar and Planetary Laboratory, University of Arizona, Tucson, AZ 85721, USA}
\affiliation{Planetary Science Institute, Tucson, AZ 85719, USA}

\author[0000-0003-3759-9080]{Tiffany Kataria}
\affiliation{Jet Propulsion Laboratory, Pasadena, CA 91011, USA}

\author[0000-0002-6694-0965]{Chao He}
\affiliation{School of Earth and Space Sciences, University of Science and Technology of China, Hefei, China}
\affiliation{Johns Hopkins University, Baltimore, MD 21218, USA}

\author[0009-0007-6910-6347]{Cara Pesciotta}
\affiliation{Johns Hopkins University, Baltimore, MD 21218, USA}

\author[0000-0003-4596-0702]{Sarah M. Hörst}
\affiliation{Johns Hopkins University, Baltimore, MD 21218, USA}

\author[0000-0001-5671-5388]{Pierre Haenecour}
\affiliation{Lunar and Planetary Laboratory, University of Arizona, Tucson, AZ 85721, USA}

\author[0000-0002-7129-3002]{Travis Barman}
\affiliation{Lunar and Planetary Laboratory, University of Arizona, Tucson, AZ 85721, USA}

\author[0000-0002-7743-3491]{Vishnu Reddy}
\affiliation{Lunar and Planetary Laboratory, University of Arizona, Tucson, AZ 85721, USA}


\author[0000-0002-8507-1304]{Nikole K. Lewis}
\affiliation{Department of Astronomy and Carl Sagan Institute, Cornell University, Ithaca, NY 14853, USA}

\author[0000-0001-7273-1898]{V{\'e}ronique Vuitton}
\affiliation{Univ. Grenoble Alpes, CNRS, IPAG, Grenoble, France}

\correspondingauthor{lhuseby@arizona.edu}
\correspondingauthor{chaohe23@ustc.edu.cn}

\begin{abstract}
Temperate sub--Neptune exoplanets could contain large inventories of water in various phases, such as water--worlds with water--rich atmospheres or even oceans. Both space--based and ground--based observations have shown that many exoplanets likely also contain photochemically--generated hazes. Haze particles are a key source of organic matter and may impact the evolution or origin of life. In addition, haze layers could provide a mechanism for lower--atmospheric shielding and ultimately atmospheric retention. Often orbiting close to M--dwarf stars, these planets receive large amounts of radiation, especially during flaring events, which may strip away their atmospheres. M--dwarf stars are known to have higher stellar activity than other types of stars, and stellar flares have the potential to accelerate atmospheric escape. In this work, we present results on laboratory investigations of UV radiation effects simulating two different stellar flare energies on laboratory--produced exoplanet hazes made under conditions analogous to water--world atmospheres. We find that both simulated flares altered the overall transmittance and reflectance of the hazes, and higher energy "flares" make those alterations more pronounced. On a larger scale, these laboratory--made hazes show potential signs of degradation over the simulated flaring period. Our results provide insight into the effects that stellar flaring events have on potential exoplanet haze composition and the ability for water--world-like exoplanets to retain their atmospheres.
\end{abstract}


\section{\textbf{Introduction}}\label{introduction}
Of the over 5500 confirmed exoplanets, the most common are considered super--Earths (1.25 -- 2.00 R$_{\rm{Earth}}$) or sub--Neptunes (2.0 -- 4.0 R$_{\rm{Earth}}$) \cite[]{Borucki2011Populations,Fressin2013OccurenceRates,Fulton2017Gap}. Among these planets, which often orbit the most common type of host star, M dwarfs (e.g., \citealt{2019HenryMdwarf}), are notable targets for further atmospheric composition characterization due to their planet--star contrast ratio and transit probability.

Recent work following the initial discovery of the radius gap \cite[]{Fulton2017Gap} has attempted to better categorize the potential compositional differences between super--Earth and sub--Neptune exoplanets. These planets have different observed mass--radius relationships, suggesting a difference in bulk densities \cite[]{Luque2022MassRadius}. Super--Earths may have silicate--iron compositions, while sub--Neptunes could have ice--silicate compositions or thin hydrogen or helium envelopes \cite[]{Mordasini2009WaterWorld,Kite2018WaterWorld}.  Some of these compositions can include water, one of the key molecules for life on Earth, as detected on TOI 270d (e.g., \citealt{Benneke2024TOI270d}). Temperate sub--Neptune exoplanets could maintain liquid water on their surfaces, creating water--worlds, or planets that have water--rich atmospheres and oceans \cite[]{Mordasini2009WaterWorld,Kite2018WaterWorld,Kite2021WaterWorld,Luque2022MassRadius,Madhusudhan2023Hycean}. 

Both space--based and ground--based observations have shown that many cooler sub--Neptune exoplanets contain clouds and hazes in their atmospheres (e.g., \citealt{Marley2013Clouds,Morley2013Clouds1214,Kreidberg20141214b,Knutson2014GJ436b,Knutson2014HD97658b,Dragomir2015GJ3470b,LibbyRoberts2020Superpuff,Kreidberg2022HD106315c}). Clouds and hazes make interpretation of exoplanet transmission spectra more difficult when they are present, causing larger than Rayleigh scattering slopes in optical wavelengths (e.g., \citealt{Ohno2020Rayleigh}), and dampening of spectral features in the infrared wavelength region, which can limit our ability to infer atmospheric compositions of these worlds (e.g., \citealt{Morley2013Clouds1214,Kreidberg20141214b,Knutson2014GJ436b,Knutson2014HD97658b,Dragomir2015GJ3470b,LibbyRoberts2020Superpuff,Kreidberg2022HD106315c}). Of particular interest are photochemical hazes, which continue to have poorly--understood growth mechanisms. Photochemical hazes are formed when gasses in an atmosphere are dissociated or ionized by high--energy radiation, undergo subsequent chemical reactions, and eventually form more chemically--irreversible, complex solids suspended in the atmosphere \cite[]{Gao2021HazeDefinition}. These hazes vary widely in composition, shape, and size, which in turn changes the optical properties of the haze, leading to variations in opacities sufficient to alter an exoplanet's spectrum \cite[]{Arney2016PaleOrangeDot,Arney2017PaleOrangeDot,Gavilan2018Haze,Corrales2023HazesCO,He2023Methods}.

The initial photochemistry necessary to trigger photochemical haze formation is likely to be present in the atmospheres of exoplanets and create thousands of organic molecules with varying compositions \cite[]{Horst2018PHAZER,He2020PHAZER,Moran2020SuperEarth,Vuitton2021}. Recent laboratory studies and modeling campaigns have suggested that hydrocarbon and organic haze particles are produced predominately through photochemistry in temperate exoplanets with equilibrium temperatures less than 1000~K \cite[]{Crossfield2017NeptuneAtmosphere,He2018MiniNeptune,Gao2020Haze,He2020PHAZER,He2020Sulfur,Yu2021Hatmosphere,Brande2024NeptuneClouds}. Given that there are no analogues for super--Earths and sub--Neptunes in the Solar System, the overall surface and atmospheric compositions are more unknown to us. Consequently, laboratory work studying how hazes form and react to irradiation provides the important empirical guidance needed to determine these compositions.  


In addition to bulk composition and clouds and hazes, another consideration in the atmospheric evolution of sub-Neptunes and super-Earths is stellar activity. Small, close--in planets can be affected by stellar activity, and more specifically, stellar flares. Flares occur when magnetic reconnection events heat localized regions of a stellar surface resulting in elevated radiation fluxes across most wavelengths \cite[]{Hurford2003GammaFlare,Benz2010FlareMechanics}. Stellar flaring impacts planets around cool stars  (T$_{\rm{eff}}$ $\sim$  3000K), where the habitable zone \cite[]{Kasting1993} is in close proximity to high--energy stellar fluxes that are potentially harmful for orbiting planets \cite[]{Howard2020FlareHab}. They also may be potentially helpful, as high-energy fluxes deliver UV radiation to the orbiting planet, which has been hypothesized to trigger biogenic processes and a potential origin of life process  \citep{Buccino2007Origins,Ranjan2017Bio,Rimmer2018Bio}. M--dwarf flares are several orders of magnitude more luminous and more frequent than those on the Sun or larger stars, releasing more high--energy particles towards the exoplanets orbiting them \cite[]{Benz2010FlareMechanics}. 
 
Stellar flares can trigger photochemistry in both gaseous and terrestrial exoplanets \cite[]{Grenfell2014,Rugheimer2015,Miguel2015FlareMN}. The increased photochemical reaction rates can alter the chemical composition and power chemical disequilibrium in an exoplanet atmosphere \cite[]{Konings2022FlaresMdwarfs}. In addition, stellar flares can drive water loss and atmospheric escape (e.g., \citealt{Lammer2007CME,Lin2019,doAmaral2022Flare,Louca2023StellarActivity}). Most sub--Neptune exoplanets have periods less than 100 days, where they are subject to high irradiation that can cause atmospheric mass loss, potentially stripping the planet from an atmosphere completely \cite[]{Mordasini2009WaterWorld}. There are competing theories on water--world atmospheric retainability. Water--worlds may be particularly susceptible to atmospheric escape, since X--ray/extreme ultraviolet (XUV) dissociates the atmospheric water vapor, and the resulting atomic hydrogen escapes through hydrodynamic winds \cite[]{Luger2015,Louca2023StellarActivity}. On the other hand, water loss on water--world exoplanet atmospheres may be suppressed by efficient water replication reactions and O$_2$ shielding of UV radiation \cite[]{Kawamura2024Waterloss}.

However, photochemical hazes may have the potential to insulate the planet's atmosphere, providing a mechanism for thermal balance as in the atmosphere of early Earth \cite[]{Jacobson2000Earth,Arney2016PaleOrangeDot,Arney2017PaleOrangeDot}. Hazes can also prevent the photolysis of biosignature gasses below in the lower atmosphere, increasing the lifetime of these gasses \cite[]{Arney2016PaleOrangeDot,Arney2017PaleOrangeDot}.
 
Critically, it remains unknown how stellar flaring affects the formation and evolution of exoplanet hazes. Previous laboratory work that subjected oxidized Titan--like hazes to UV radiation revealed the formation of oxygenated bonds and new electronic transitions \cite[]{Gavilan2018Haze,Carrasco2018TitanUV}. Since the optical properties of organic hazes are needed to refine exoplanet atmospheric modeling, temperate sub--Neptune exoplanet haze molecular properties and changes over time are necessary to understand the spectral variations and persistence of such particles as a function of irradiation. 
 
The aim of this work is to quantify spectral changes to exoplanet hazes as they are exposed to UV radiation simulating a higher energy event like stellar flaring, as M--dwarf flares occur on relatively short timescales. This work will help us begin to understand if water--world exoplanets would be able retain their haze layer and overall atmosphere in the presence of such stellar activity. We therefore subjected two laboratory--made “water--world” haze samples to varying UV irradiation environments to assess how the hazes evolve over time. We measured the transmittance and reflectance spectra of the hazes before and after UV irradiation across a broad wavelength range (from Far Ultraviolet (FUV) to mid--IR, 0.2 -- 9 $\mu$m), which overlaps with both HST, JWST, and the upcoming Habitable Worlds Observatory (HWO) instruments accessible for transiting exoplanet observations. This work will improve our understanding of haze evolution in sub--Neptune exoplanet atmospheres and our understanding of how stellar flares can impact the composition and atmospheric chemistry at work in these exotic worlds.

In Section \ref{Methods}, we describe our experimental methods including the haze sample production (\ref{hazeproduction}) and our haze UV bombardment process (\ref{FTIR}). Section \ref{Results} describes the findings of our experimental work, including the physical appearance of our samples (\ref{hazeimages}) in addition to transmittance (\ref{Transresults}), reflectance (\ref{Refresults}), and time--series results (\ref{TDresults}). In Sections \ref{sec:Discussion} and \ref{conclusion}, we discuss the implications of our results and their importance to the wider exoplanet science community. 

\section{\textbf{Experimental Methods}} \label{Methods}
\subsection{Haze Sample Production}\label{hazeproduction}
\begin{figure*}[ht]
\centering
\includegraphics[width=\textwidth]{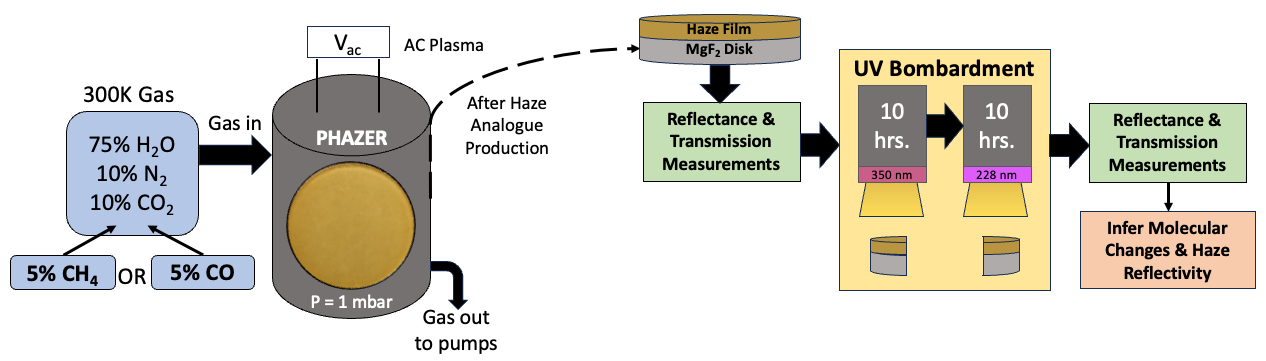}
\caption{Streamlined schematic of the experimental setup, simulated atmospheric compositions and conditions, UV bombardment process, measurements, and experimental outcomes. Two laboratory hazes were produced (the initial conditions varying only in the minor carbon source) in the PHAZER chamber \citep{He2017PHAZER} by exposing the gas mixture at room temperature to an AC plasma source. After the hazes were produced, each half of the films were exposed to UV radiation under two different filters. The transmittance and reflectance spectra pre-- and post--UV irradiation of both samples were measured using a FTIR spectrometer to quantify molecular changes and destruction during the irradiation process.}
\label{fig:ProductionSchematic}
\end{figure*} 


Detailed compositions of cool, high mean molecular weight exoplanet hazes are not well measured due to the unknown mechanism in which they are formed. Therefore, we must rely on chemical equilibrium calculations in which inputs of exoplanet temperature, pressure, and metallicity are used to guide the atmospheric composition \cite[]{Moses2013Chemistry}. Our atmospheres are a simplification of that work, which consists of a background gas of water, mixed with other molecules found in observations of higher metallicity atmospheres at a mixing ratio high enough to participate in photochemistry. These compositions provided a starting point for our experiment investigating the irradiation of photochemically--produced hazes in the atmospheres of sub--Neptunes. 

The physical hazes in this work are made to mimic a potential sub--Neptune water--world atmosphere. The water--world compositions we use here are informed by 1000$\times$ solar metallicity abundances, or increased atmospheric abundances of elements heavier than hydrogen and helium, as motivated by previous laboratory experiments \cite[]{He2018HazeRough,He2018MiniNeptune,Horst2018PHAZER}. These past experiments using 1000x metallicity found that laboratory hazes generated from high abundances of water vapor had high production rates at 300K (e.g., \citealt{Horst2018PHAZER}). In addition, both CH$_4$-- and CO--containing atmospheres, also found in previous 1000x metallicity experiments, increased haze production rates (e.g., \citealt{Horst2014HazeCO,Horst2018PHAZER}). In high metallicity atmospheres, CO$_2$ and N$_2$ \cite[]{Deming2017ExoAtmo} can be very abundant and are added to our compositions in equal amounts (10\%). Then, either CO or CH$_4$ is added at the 5\% level to ensure that they will interact photochemically in an observable manner \cite[]{Morley2013Clouds1214}. This work will also help investigate the role that the minor carbon source has in organic haze formation and respective resistance to simulated stellar flaring events. 

\begin{table}[h!]
\centering
\begin{tabular}{lcc}
\toprule
&\textbf{Sample 1} & \textbf{Sample 2} \\ 
\midrule
& 75\% H$_{2}$O            & 75\% H$_{2}$O    \\
& 10\% N$_{2}$             & 10\% N$_{2}$            \\
& 10\% CO$_{2}$            & 10\% CO$_{2}$            \\
& 5\% CO              & 5\% CH$_{4}$       
       \\ \bottomrule
\end{tabular}
\caption{Gas compositions by mixing ratio of the two samples made at 300K, 667 Pa. Gas mixtures represent approximately equilibrium compositions at 1000x Solar metallicity, scaled slightly for experimental simplicity. The two haze samples differ in the minor carbon source.} 
\label{table:GasComp}
\end{table}

Table \ref{table:GasComp} breaks down the initial gas compositions used to create the two haze samples for this experiment. Two haze samples analogous to sub--Neptune water--world atmospheres were made from H$_2$O--dominated atmospheric mixes. These samples were produced at the Johns Hopkins University Planetary Haze Research (PHAZER) laboratory \cite[]{He2017PHAZER}. Here, we briefly describe the haze production process, also further described in \citet[]{Horst2018PHAZER, He2017PHAZER,He2020PHAZER,He2023Methods}.

Figure~\ref{fig:ProductionSchematic} illustrates our laboratory setup. All gases in Table~\ref{table:GasComp}, except for water vapor, were premixed into a stainless--steel cylinder with high--purity gases purchased from Airgas (N$_{2}$ -- 99.9997\%, CO$_{2}$ -- 99.999\%, CH$_{4}$ -- 99.999\%, CO -- 99.999\%). The gas mixture flowed at a rate of 2.5 standard cubic centimeters per minute (sccm). The haze analogues were produced at a total pressure of 667 Pa. Water vapor (HPLC water, Fisher Chemical) is introduced to the system at a pressure of 500 Pa (667Pa $\times$ 75\% = 500Pa) from water ice at 270 K and allowed to equilibrate in the PHAZER chamber. The gas mixture was then exposed to an AC glow discharge in the PHAZER reaction chamber (Figure \ref{fig:ProductionSchematic}). The AC glow discharge is not analogous to any planetary atmospheric mechanism, like lightning, or stellar phenomenon, like a coronal mass ejection, but is rather representative of generic energetic processes happening in the upper atmosphere, allowing us to have a higher production rate than other methods (See \citealt{Horst2018PHAZER} and \citealt{Cable2012Lamps} for further discussion). By utilizing this method, the AC glow discharge is able to dissociate stable molecules without altering the ambient gas temperature found in the chamber. This discharges electrons into the gas composition flow, which initiates the chemical processes necessary to form solid particles and new gas--phase products. We note that the power output used in the AC plasma source \cite[]{He2020PHAZER, He2020Sulfur} was 170 W m$^{-2}$, significantly higher than the output flux received by temperate exoplanets orbiting M dwarf stars in quiescence (e.g., GJ 1214b $\sim$3.5 W m$^{-2}$). This is due to the fact that laboratory simulations usually use a higher energy density to accelerate the chemical process to analyze and observe the chemical impact within a reasonable timeline (e.g., \citealt{He2023Methods}). 

To prepare the film sample for this experiment, an optical--grade MgF$_2$ substrate plate (diameter: 25 mm, thickness: 1 mm, Crystran) was placed into the PHAZER chamber for sample collection. MgF$_2$ substrate plates were utilized as MgF$_2$ is chemically--inert, transparent in the wavelength range of 0.1 -- 9 $\mu$m, and highly resistant to UV degradation, so that it would not react with our sample during irradiation. The newly--formed solid particles settled down the chamber onto our MgF$_2$ substrates as thin films. The AC discharge exposure continued for 72 hours. The gas flow was then turned off and samples were then kept under vacuum for 48 hours to remove any volatile components. 

After the vacuum purge and return to ambient pressure, the PHAZER chamber was transferred to a dry (<0.1 ppm H$_{2}$O), oxygen free (<0.1 ppm O$_{2}$), N$_{2}$ glove box (purchased from Inert Technology Inc.) at Johns Hopkins University to collect the haze products with no exposure to the Earth's atmosphere. The haze sample films were transferred to plastic cases, which were sealed with parafilm and covered with aluminum foil for storage. The samples were kept stable for 8 months protected from Earth ambient atmosphere and light sources before being transported, still sealed and covered, to the University of Arizona for UV irradiation experiments. This storage period has previously been shown not to affect PHAZER samples, with particles maintaining their original compositions \cite[]{Moran2022Triton}. 

\subsection{FTIR Spectroscopy Measurements and Calculations}\label{FTIR}

\subsubsection{FTIR Measurements and UV Bombardment}

For irradiation and measurement (University of Arizona, Tucson, AZ), we placed the samples in a room--temperature (294K)  N$_2$--purged sample chamber (99.999\%, AirProducts) in the Fourier Transform Infrared (FTIR) spectrometer (Nicolet iS 50R Benchtop FTIR) on a sliding stage and custom sample holder that contained two standards and the sample. A fiber optic cable bundle was fed through the chamber. These cables delivered near--UV light from a deuterium lamp (Aventes Avalight DH--S--BAL Balanced Deuterium Lamp, Power: 78 W, Wavelength Range: 200--600 nm), and one cable collected reflected light and sent it to the UV-VIS (0.2 -- 0.6 $\mu$m) spectrometer (Avantes SensLine AvaSpec-ULS2048LTEC). We took visible to thermal infrared (0.37 -- 9 $\mu$m) reflectance spectra at 0$^{\circ}$ incidence periodically with both the FTIR and the UV--VIS spectrometer. We used two reflectance standards for different wavelength ranges: a Spectralon polytetrafluoroethylene (PTFE) standard for 0.2 -- 1.16 $\mu$m and diffuse gold for the 1.16 -- 9 $\mu$m range. The sliding stage and custom sample holder enabled standards and samples to be moved in and out of measurement and UV irradiation positions of the two spectrometers and light source without breaking N$_2$ purge and to better focus the UV light onto the samples. At the beginning and end of each irradiation run, we changed the sampling configuration from reflectance to transmittance, and transmittance spectra were taken with just the FTIR. We compared all transmittance measurements against a blank MgF$_2$ substrate plate identical to the haze product sample plate.

\begin{table} [h!]
    \centering 
    \begin{tabular}{llll}
    \midrule
    \multicolumn{1}{p{2cm}}{\centering \textbf{Wavelength \\ (${\mu}$m)}}
    & \multicolumn{1}{p{2.2cm}}{\centering \textbf{Measurement}}
    & \multicolumn{1}{p{2.5cm}}{\centering \textbf{Detector, \\ Beamsplitter, lamp}}
    \\ \midrule
    0.2 -- 0.6  & Reflectance & \multirow{2}{10em}{Silicon CCD, None, Deuterium} \\
    \\
    0.37 -- 1.16 & \multirow{1}{5em}{Reflectance, Transmittance} & \multirow{2}{10em}{Si--diode, Quartz, Quartz -- tungsten halogen}\\ \\ \\
    0.9 -- 2.6 & \multirow{1}{5em}{Reflectance, Transmittance} & \multirow{2}{10em}{MCT, Quartz, Quartz -- tungsten halogen}\\ \\
    1.25 -- 9 & \multirow{1}{5em}{Reflectance, Transmittance} & \multirow{2}{10em}{MCT, KBr, Silicon -- nitride globar} \\ \\
    \midrule
    \end{tabular}
    \caption{Breakdown of instruments used by wavelength for each measurement technique during the experiment}
    \label{table:Instruments}
\end{table}

As seen in Table \ref{table:Instruments}, for reflectance from 0.2 to 0.6 $\mu$m we used an Avantes spectrometer, with back thinned silicon CCD detector using a deuterium lamp as the light source. Measurements were taken throughout the bombardment process (described below) every 1.5 hours using the Avantes spectrometer with 5 scans for a 1.4 nm FWHM resolution to monitor compositional changes over time. We conducted measurements with the FTIR at 0$^{\circ}$ incidence with a maximum resolution of 0.09 cm$^{-1}$ in 3 wavelength ranges to provide the best signal from various detectors, beam splitters, light source and reflectance standards. From 0.37 to 1.16 $\mu$m in both reflectance and transmittance, we took 200 scans with a Si--diode detector and quartz beamsplitter with quartz–tungsten halogen lamp as a light source. From 0.9 to 2.6 $\mu$m in reflectance and transmittance, we took 200 scans with a Mercury Cadmium Telluride (MCT) detector and quartz beamsplitter and a quartz–tungsten halogen lamp. From 1.25 to 9 $\mu$m in reflectance and transmittance, we took 200 scans with a MCT detector and potassium bromide (KBr) beamsplitter and a silicon-nitride globar light source. Overlap between detectors allowed for calibration between wavelength regions.

We then irradiated the samples with two different bandpasses of light (215 -- 245 nm, peak: 228 nm, 20\% throughput, Andover Corporation \footnote{https://www.andovercorp.com/search/228FS25-25}; and 320 -- 380 nm, peak: 350 nm, 80\% throughput, Baader Planetarium \footnote{ https://www.astro-physics.com/bpu2}) for 10 hours per filter to simulate the effect of M dwarf host star flaring on an exoplanet atmospheric haze. The most active M dwarf stars produce dozens of flares per day along with higher energy "super flares" approximately once per month \cite[]{Loyd2016Flare}. Flares on these planets can last from tens of minutes to hours \cite[]{Brasseur2019} and therefore 10 hours was chosen as a reasonable approximation given the variance of flare energies, timescales, and experimental capabilities. See further discussion in Section \ref{sec:Discussion} for how this laboratory energy density and flaring period compares to typical M dwarfs. 

One half of the sample was bombarded with UV light through the 350 nm filter, and then subsequent transmittance and reflectance measurements were taken. We rotated the sample using the sliding stage and sample holder and bombarded the other half with a UV lamp through the 228 nm filter with the same measurements taken as the other filter. This ensured that there would be no cumulative irradiation effects on the sample, as one half received radiation through the 350 nm filter, and the other half through the 228 nm filter. The output of our UV light during the irradiation period is 1.1 W/m$^{2}$. The filters had different peak wavelength throughputs to compare a representative higher and lower energy flare impacting the atmosphere of the planet in the UV region. After each irradiation filter, we took transmittance and reflectance measurements again to observe any changes due to the UV bombardment. 

\subsubsection{Irradiated Sample Post--Processing}
For the UV-Vis data in transmittance and reflectance, clear outliers were removed (e.g. 484.23 nm -- 487.15 nm, 654.66 nm -- 656.95 nm). These points were demarcated as outliers as their values were $\sim$ 10$^5$ higher than the rest of the sample data. After the removal of the clear outliers, interference fringes were observed between 35000 -- 18000 cm$^{-1}$ (0.28 -- 0.53 $\mu$m) in the time--dependent visible wavelength data. These interference fringes occur only in optical wavelengths due to multiple reflections between different film thicknesses across the sample \cite[]{1987NeriInterference}. 

For a thin film, as seen in these experiments, we can use interference fringes to calculate a film thickness ($t$). We follow the process from \citealt{He2022Titan} using Equation (\ref{eq:Film thickness equation}) \cite[]{Rancourt1987OpticalTF,Stenzel2005Fringes}:

\begin{align}
    \label{eq:Film thickness equation} 
    t = \frac{1}{2 \sqrt{n^2 - \sin^2i}}  \times  \frac{x}{(\nu_1 - \nu_2)}.
\end{align}

Here, $n$ is the refractive index of our material, $i$ is the angle of incidence, and $x$ is the number of fringes between two specified wavenumbers, $\nu_1$ and $\nu_2$. From the observed fringes seen in Figure \ref{fig:CH4_320} and \ref{fig:CH4_240}, we can get $x$, $\nu_1$, and $\nu_2$. The angle of incidence in our experiment was normal at 0$^{\circ}$. Using an assumed $n$ value of 1.7 from previous laboratory experiments (e.g., \citealt{He2022Titan}), we then calculate the film thickness pre-- and post--irradiation. Our results are shown in Table \ref{table:Thickness}.

After calculating the film thickness, we applied the correction in Equation (\ref{eq:Interference equation}) following the moving average method of \cite{1987NeriInterference} as used in \cite{He2022Titan} and \cite{Moran2022Triton} to eliminate the effect of the fringes on our resulting spectra. This procedure normalized the absorbed power by the transmitted power with assumptions that 1) the processed spectrum has equally--spaced data points, and 2) that the fringes have similar amplitudes. This correction follows the form:

\begin{align}
    \label{eq:Interference equation} 
    F(X_n) = \frac{2G(X_n) + G(X_{n+m}) + G(X_{n-m})}{4}.
\end{align}

Here, $X_n$ is the $n$th abscissa, $F(X_n)$ is the fringe-removed spectrum value at $X_n$, $G(X_n)$ is the original spectrum value at $X_n$, $G(X_{n+m})$ and $G(X_{n-m})$ are the original spectrum values at shifted abscissae and 2$m$ is the maximum integer number of points contained in the interval $d$, which is the average fringe spacing. For our transmittance and reflectance spectra, the average fringe spacing ($d$) was approximately 2700 -- 3000 cm$^{-1}$ depending on sample and there are 1336 -- 1500 points ($m =$ 678 -- 728) contained in the interval $d$. The reflectance and transmittance spectra in the near--infrared and thermal--infrared were not treated because they are unaffected by fringing. Once the removal of the interference fringes was completed, we then compared the spectra between pre-- and post--UV irradiation across both filters. 

\section{\textbf{Results}} \label{Results}
\subsection{Haze Sample Films} \label{hazeimages}
\begin{figure}[H]
    \centering
    \includegraphics[width=0.7\linewidth]{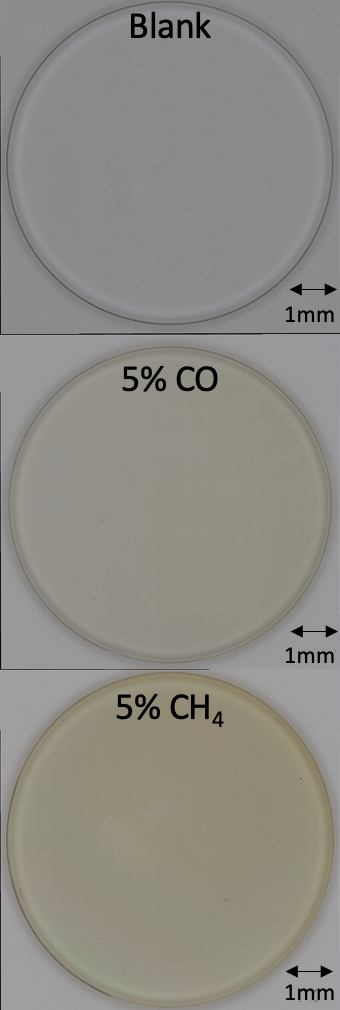}
    \caption{Optical reflected light mosaic images (Keyence Digital Microscope, University of Arizona) of the post--UV bombardment water--world hazes deposited onto MgF$_{2}$ substrate disks. The image labeled blank is of a clean MgF$_{2}$ substrate disk. The image labeled 5\% CH$_4$--derived haze sample is more yellow in color than the image labeled 5\% CO--derived haze sample, indicating the differing composition and haze--forming efficiency of the samples.}
    \label{fig:HazePictures}
\end{figure}

Images of our experimental haze samples post-UV bombardment show differences in the physical haze color between the two samples. Figure \ref{fig:HazePictures} shows optical reflected light mosaic images of the haze samples and one of a MgF$_{2}$ blank as reference. The haze samples are yellow/orange in color, with the 5\% CH$_4$--derived haze sample being slightly darker than the 5\% CO--derived haze sample. Previous studies (e.g., \citealt{Horst2018PHAZER,He2018MiniNeptune,He2018HazeRough}) have shown that darker colors in similar temperature hazes are due to haze production rate differences in addition to compositional changes. This could mean that the 5\% CH$_4$ sample had a higher production rate as compared to the 5\% CO sample. More work is necessary to determine the complex chemical processes that produce the resulting color changes, and the implications they have for photochemical haze production mechanisms. However, this is outside the scope of this study and will be explored in complementary studies by a subset of our team. Pre--irradiation images are unavailable to limit atmospheric contamination with the haze samples, as the imager was not under N$_2$ purge. During irradiation, no physical color changes are seen by eye. Both haze samples have a smooth surface, with a post--irradiation calculated root mean square (RMS) surface roughness of 17.67 nm for the 5\% CO--derived haze sample and 3.08 nm for the 5\% CH$_4$--derived haze sample. 

We also calculated the film thickness of the 5\% CH$_4$--derived haze sample both pre-and post-irradiation (Table \ref{table:Thickness}). We compute only one value for the film thickness both pre- and post-irradiation of the 5\% CO--derived haze sample, as there was no observed change in reflectance, as seen in Figure \ref{fig:CO-UVTotal}. Between pre-- and post--irradiation in the 5\% CH$_4$ derived haze sample, there is a loss in thickness across both filters from 0.911 ${\mu}$m 

to 0.862 ${\mu}$m and to 0.891 ${\mu}$m respectively (Table \ref{table:Thickness}). The haze film is thinner post--irradiation across the 350 nm filter in comparison to the 228 nm filter. However, we note that error in the thickness of the sample after UV-bombardment with both the 228 nm filter and the 350 nm filter overlaps with the pre--irradiation sample thickness. Therefore we cannot conclude if higher--energy flares make hazes thinner than lower--energy flares, or if the changes in thickness are significant between pre-- and post--irradiation within the limits of our experimental uncertainty.

\begin{table} [h!]
    \centering 
    \begin{tabular}{lc}
    \midrule
    \multicolumn{1}{p{2cm}}{\centering \textbf{Sample}}
    & \multicolumn{1}{p{3cm}}{\centering \textbf{Thickness  \\ (${\mu}$m)}}
    \\ \midrule
    \multirow{1}{15em}{5\% CH$_4$: Pre--Irradiation}  & 0.911 $\pm$ .094 \\ \\
    \multirow{1}{12em}{5\% CH$_4$: Post--Irradiation (350 nm)} & 0.862 $\pm$ .021\\ \\
    \multirow{1}{12em}{5\% CH$_4$: Post--Irradiation (228 nm)} & 0.891 $\pm$ .056\\ \\
    \midrule \midrule
     \multirow{1}{12em}{5\% CO: Pre-- and Post--Irradiation}  & 0.368 $\pm$ .037 \\ \\
     \midrule
    \end{tabular}
    \caption{Calculated film thicknesses of the haze samples based upon observed optical fringes (see Equation \ref{eq:Film thickness equation}). The 5\% CO derived haze sample is thinner than the 5\% CH$_4$ derived haze sample, consistent with their appearances in Figure \ref{fig:HazePictures}.}
    \label{table:Thickness}
\end{table}

\subsection{Transmittance Spectra Results} \label{Transresults}

\begin{figure}[h!] 
    \centering
    \includegraphics[angle=-90,width=\columnwidth]{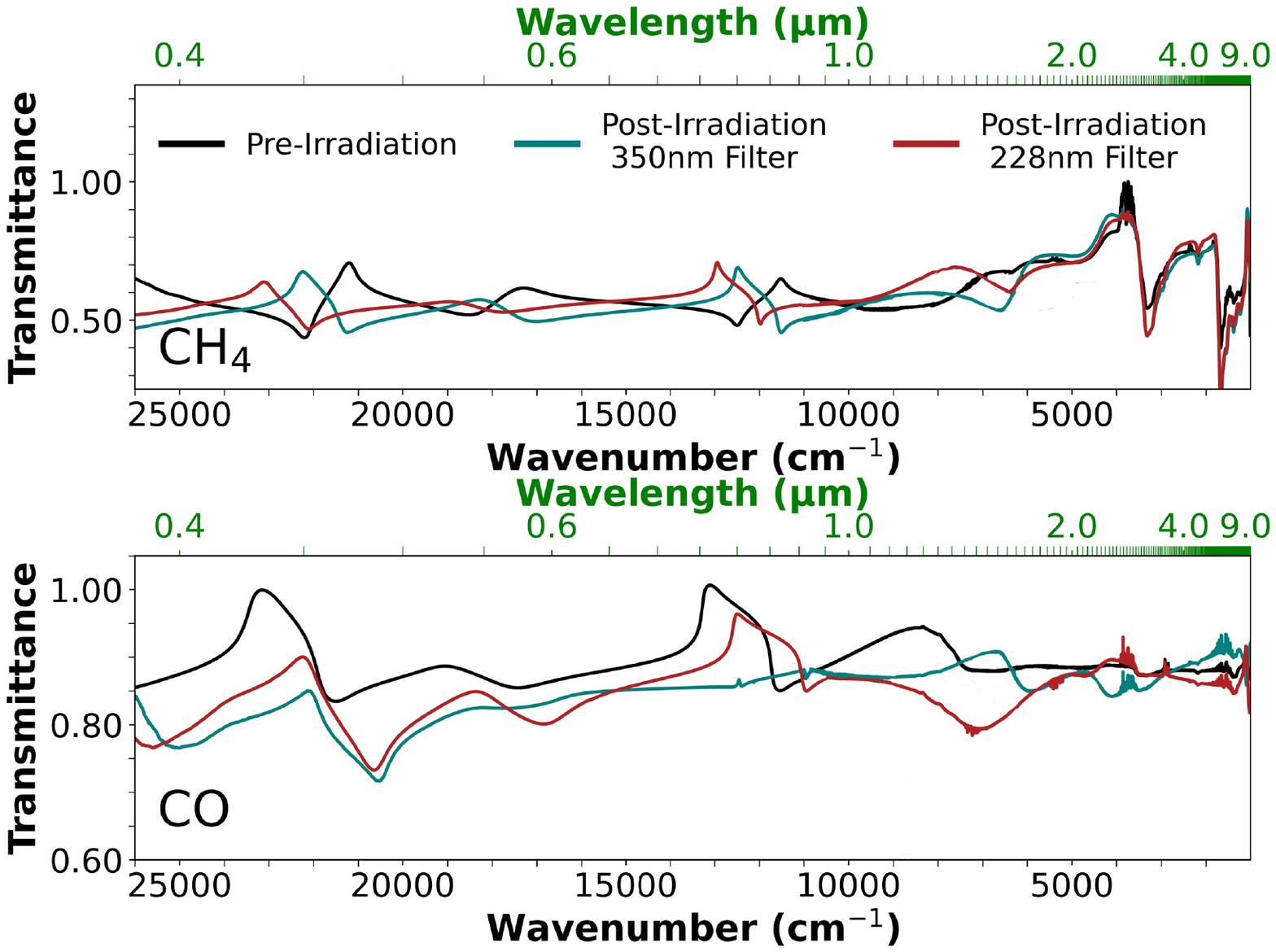}
    \caption{Transmittance spectra of our 5\% CH$_4$ atmosphere haze sample (top) and 5\% CO atmosphere haze sample (bottom) in the visible to mid--IR wavelength region (26000 -- 1100 cm$^{-1}$, 0.4 -- 9 $\mu$m). The black line represents the samples before the irradiation process began, and the teal and red lines represent the post--irradiation spectrum using the 350 nm and 228 nm filters respectively.} 
    \label{fig:TotalTransmission}
\end{figure}

Exoplanet radiative transfer models require haze opacities to accurately understand their impact on exoplanet transmission spectra. Haze opacities can be calculated from the refractive indices of haze samples, derived from transmittance and reflectance spectra, which we measure here. Figure \ref{fig:TotalTransmission} shows the transmittance spectra of both samples pre-- and post--irradiation. The spectral feature determinations below are made following the IR spectrum tables from Sigma Aldrich in addition to previous laboratory work \cite[]{He2022Titan,He2023Methods}.

\begin{figure}[h] 
    \centering
    \includegraphics[angle=-90,width=\columnwidth]{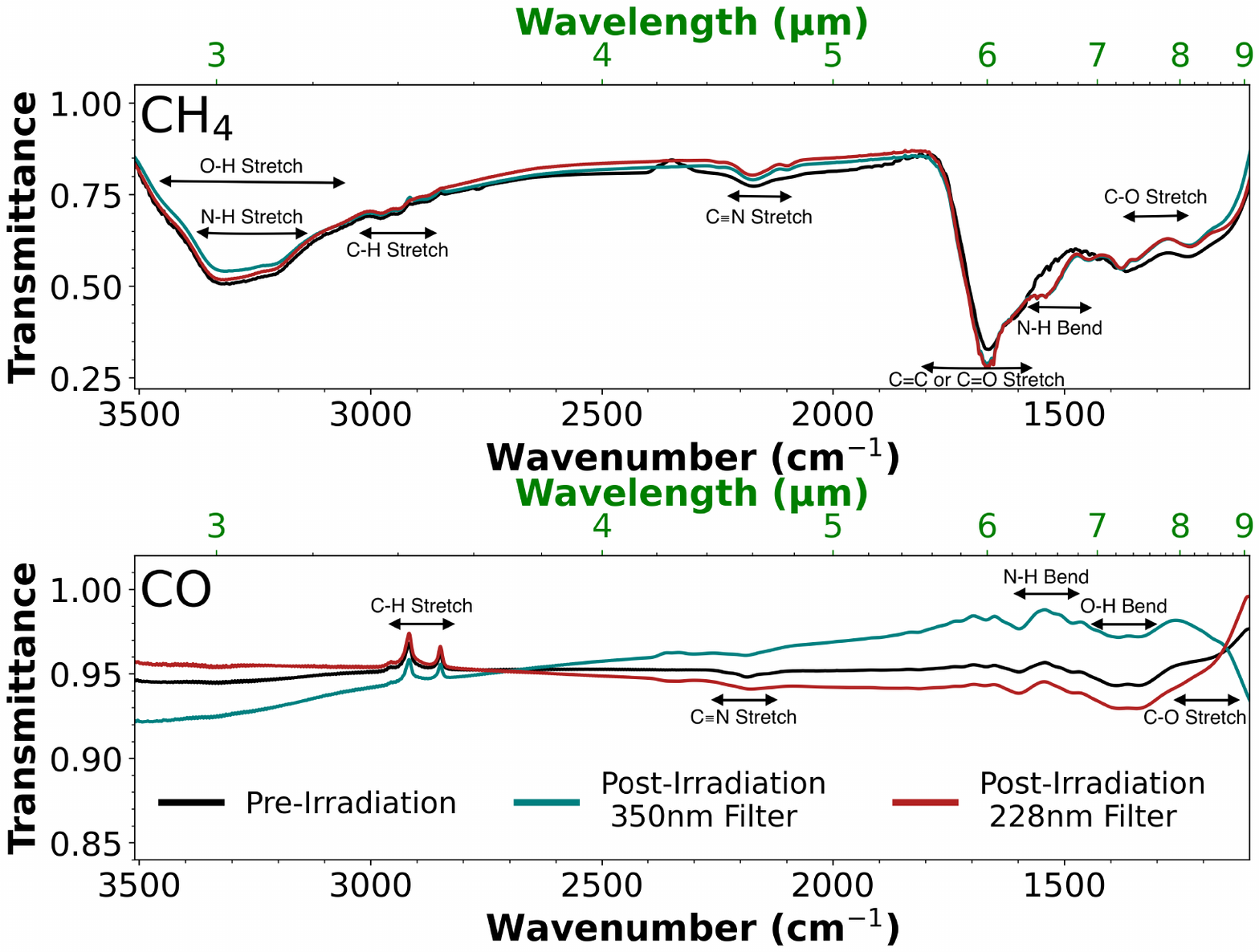}
    \caption{Enlarged spectrum between 3500 -- 1100 cm$^{-1}$ (2.5 -- 9 $\mu$m) of the 5\% CH$_4$ (top) and 5\% CO (bottom) atmosphere haze samples as a function of wavelength in transmittance with major spectral features labeled.}
    \label{fig:TotalTrans}
\end{figure}

We stitch together three different spectra from 0.37--1.16~$\mu$m, 0.9--2.6 $\mu$m, and 1.25--9 $\mu$m to generate the full haze sample continuum in transmittance. As seen in Figure \ref{fig:TotalTransmission}, the general spectral shape and features of both the CH$_4$ and CO spectra are similar throughout the irradiation process. There are two larger absorption features at 0.44 $\mu$m and 0.8 $\mu$m in both samples. This suggests that there may be larger aromatic compounds or other organic bonds present in the haze samples \cite[]{Krevelen2009OpticalProp}. The spectrum is relatively featureless from 0.8--2.5 $\mu$m. Larger spectral features that can be distinguished are found from 2.5--9 $\mu$m as is typical of organic material \cite[]{He2018MiniNeptune}. In Figure \ref{fig:TotalTrans} we narrow the wavelength to 3500--1100 cm$^{-1}$ (2.5--9 $\mu$m) for closer inspection of spectral features indicative of particular molecular functional groups.

\subsubsection{CH$_4$--Derived Haze Sample}
 For the 5\% CH$_4$ atmosphere haze sample in transmittance, Table \ref{table:CH4 Trans} and Figure \ref{fig:TotalTrans}, top panel, identifies the bonds responsible for each spectral feature. In Tables \ref{table:CH4 Trans}, \ref{table:CO Trans}, and throughout the results, relative change is defined as the difference between the post-- and pre--irradiation spectrum with respect to the pre--irradiation spectrum of each filter. There are characteristic bonds of O--H, C--H, C=O, C--O, and N--H. The spectra above 9 $\mu$m could be affected by the MgF$_2$ substrate disk, as the transmittance of the disk at those wavelengths is near zero, so small differences including noise may cause large spectral differences. We exclude these longer wavelengths from our analysis. The large absorption feature between 3350--3200 cm$^{-1}$ (2.97--3.12 $\mu$m; Figure~\ref{fig:TotalTrans}, top panel) is largely due to O--H stretching, indicative of intramolecular bonded alcohols in this haze sample, in addition to N--H amine stretching. Both filters during the irradiation process slightly increase the overall transmittance of the sample over this absorption feature, meaning that there are lower amounts of O--H bonds present on the film in that spectral feature wavelength range. One potential destruction mechanism could be that O--H bonds within the CH$_4$--derived haze sample have dissociated and recombined with other molecules to create more complex alcohols at a different wavelength, or dissociated completely, leading to an increase in transmittance found in the transmittance spectra. 

\begin{figure}[h!]
    \centering
    \includegraphics[width=\columnwidth]{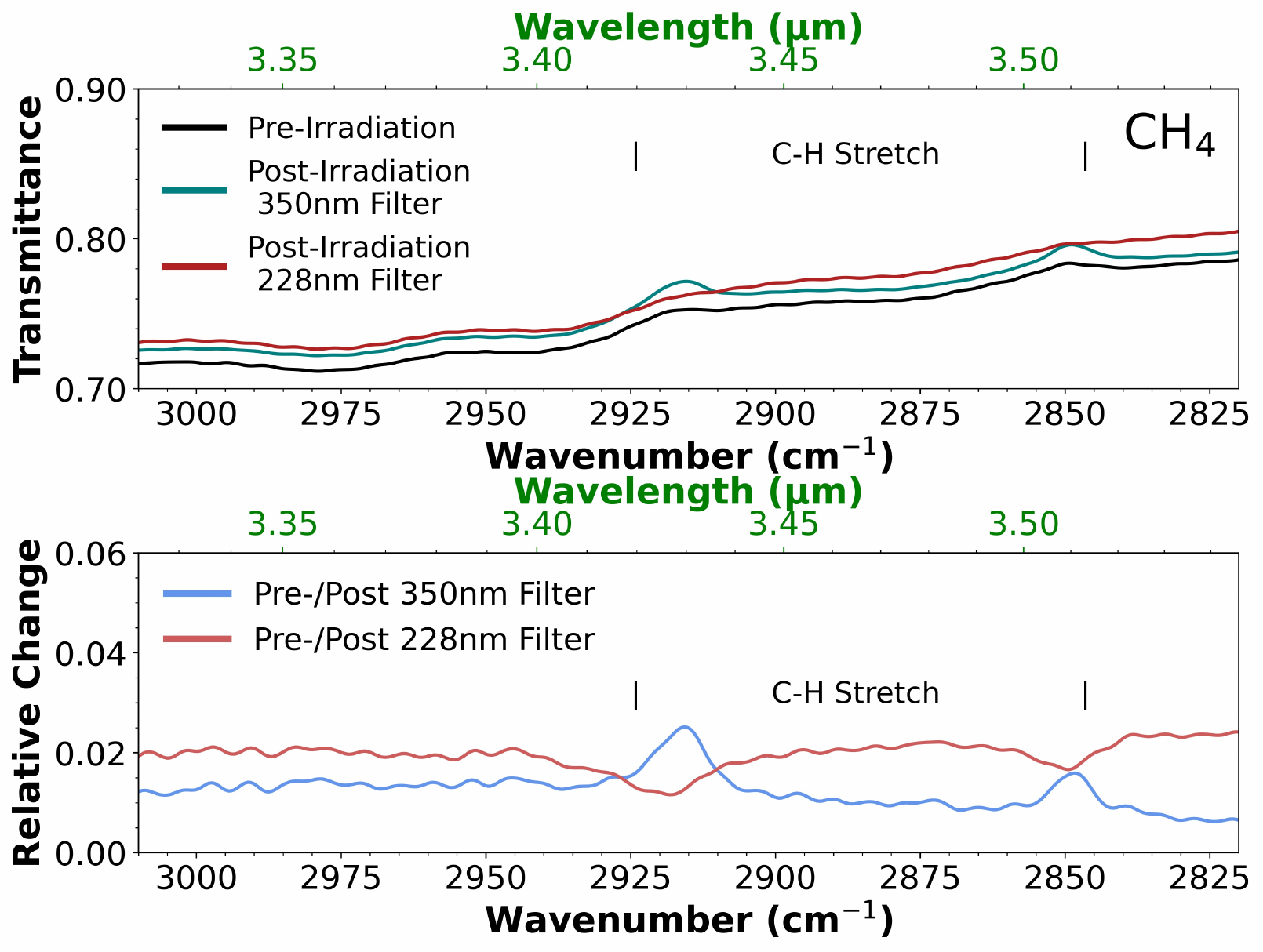}
    \caption{Top: the C--H stretching mode of an alkane spectral feature of the 5\% CH$_4$ atmosphere--derived sample pre-- and post--UV irradiation in transmittance. The post--irradiation through the 350 nm filter had larger increases in transmittance and created "bumps" in the spectra. Bottom: the relative change between the pre--irradiation and post--irradiation spectra of each filter respectively due to the UV bombardment.}
    \label{fig:Trans3000}    
\end{figure}

There are two different peaks at 2925 cm$^{-1}$ (3.42 $\mu$m) and 2850 cm$^{-1}$ (3.51 $\mu$m), both indicating the presence of C-H stretching due to alkanes present in the haze sample (Figure~\ref{fig:Trans3000}, top panel). During the irradiation process, both features again show a slight increase in transmittance. This also occurs in the 5\% CO atmosphere haze sample (Figure~\ref{fig:TotalTrans}, bottom panel) , although the 5\% CH$_4$ atmosphere haze sample has larger continuum and spectral feature changes during irradiation (Figure~\ref{fig:Trans3000}). More specifically, the spectra across the 350 nm filter not only shows a slight increase in overall transmittance but has a slightly more defined doublet absorption feature as compared to both the pre--irradiation spectrum and the spectra across the 228 nm filter (Figure~\ref{fig:Trans3000}). In the spectrum across the 228 nm filter, the doublet feature is not apparent, meaning that the higher--energy radiation may have eliminated the spectral feature that occurs due to the C--H stretching. This potentially means that the large amounts of C--H bonds in the haze sample may have been fully dissociated due to the higher--energy radiation (Figure~\ref{fig:Trans3000}).  

\begin{table*}[ht]
\centering
\caption{Spectral features and corresponding functional groups identified in the 5\% CH$_4$--derived atmospheric haze sample 
pre-- and post--irradiation in transmittance and reflectance. Spectral features with an * are seen in both spectra.} 
\textbf{\large Transmittance}
\begin{tabular*}{\linewidth}{lllll}
\midrule
\multicolumn{1}{p{2cm}}{\centering \textbf{\large Frequency \\ (cm$^{-1}$)}}
& \multicolumn{1}{p{2.3cm}}{\centering \textbf{\large Wavelength \\ (${\mu}$m)}}
& \multicolumn{1}{p{3.8cm}}{\centering \textbf{\large Functional Group \\ (bond, type)}}
& \multicolumn{1}{p{2.5cm}}{\centering \textbf{\large Intensity}}
& \multicolumn{1}{p{3.7cm}}{\centering \textbf{\large Relative Change \\ (228 nm, 350 nm)}}
\\ \midrule
3350 -- 3200  & 2.97 -- 3.12  & *O--H alcohol (intermolecular bonded) & strong, broad & no change, $\boldsymbol{\uparrow}$ 0.08 \\
3400 -- 3225  & 2.95 -- 3.10  & *N--H stretching (amine group) & medium, broad &  no change, $\boldsymbol{\uparrow}$ 0.08 \\
2925 -- 2850  & 3.42 -- 3.51  & *C--H stretching (alkane) & medium, doublet & no change, $\boldsymbol{\uparrow}$ 0.02\\
2175  & 4.63  & C$\equiv$N stretching (conjugated) & strong, broad & $\boldsymbol{\uparrow}$ 0.05, $\boldsymbol{\uparrow}$ 0.03\\
1800 -- 1500  & 5.50 -- 6.50  & C=C / C=O stretching & strong, broad &  $\boldsymbol{\uparrow}$ 0.41, $\boldsymbol{\uparrow}$ 0.41\\
1620 -- 1580  & 6.2 -- 6.35  & N--H bending (amine) & medium, broad & $\boldsymbol{\uparrow}$ 0.07, $\boldsymbol{\uparrow}$ 0.07\\
1126 -- 1104  &  8.88 -- 9.06  & C--O stretching & strong, shoulder & $\boldsymbol{\uparrow}$ 0.04, $\boldsymbol{\uparrow}$ 0.12\\
\midrule
\\ \multicolumn{5}{c}{\textbf{\large Reflectance}} \\
\midrule

3500 -- 2700  & 2.86 -- 4.0  & *O--H stretching (alcohol/carbonic acid) & strong, broad & $\boldsymbol{\downarrow}$ 0.28 , $\boldsymbol{\downarrow}$ 0.3\\
3400 -- 3225  & 2.95 -- 3.10  & *N--H stretching (amine group) & medium, broad & $\boldsymbol{\downarrow}$ 0.28,  $\boldsymbol{\downarrow}$ 0.3 \\
3075 -- 3025  & 3.25 -- 3.30  & *C--H stretching & medium, broad & $\boldsymbol{\downarrow}$ 0.3,   $\boldsymbol{\downarrow}$ 0.3\\
2740 -- 2700 & 3.66 -- 3.70  & C--H stretching (aldehyde, doublet) & medium, sharp & $\boldsymbol{\downarrow}$ 0.27, $\boldsymbol{\downarrow}$ 0.25\\
2260 -- 2222  & 4.42 -- 4.5  & C$\equiv$N stretching (nitrile group) & weak, broad & $\boldsymbol{\downarrow}$ 0.3, $\boldsymbol{\downarrow}$ 0.3\\
1740 -- 1700  & 5.75 -- 5.87  & C=O stretching (aldehyde) & strong, broad & $\boldsymbol{\downarrow}$ 0.36, $\boldsymbol{\downarrow}$ 0.36\\
1675 -- 1640 & 5.96 -- 6.06  & C=O, C=N, or C=C stretching & weak, broad & $\boldsymbol{\downarrow}$ 0.22, $\boldsymbol{\downarrow}$ 0.22\\
1425 -- 1375 & 7.0 -- 7.25  & C--H bending (aldehyde) & medium, sharp & $\boldsymbol{\downarrow}$ 0.42, $\boldsymbol{\downarrow}$ 0.4\\
1275 -- 1220  & 7.85 -- 8.17  & C--O stretching (ether group) & strong, broad & $\boldsymbol{\downarrow}$ 0.44, $\boldsymbol{\downarrow}$ 0.4\\
1170 -- 1130 & 8.52 -- 8.85  & C--O stretching (tertiary alcohol) & weak, sharp & $\boldsymbol{\downarrow}$ 0.54, $\boldsymbol{\downarrow}$ 0.52\\
1085 -- 1050 & 9.28 -- 9.53  & C--O stretching (primary alcohol) & strong, broad & $\boldsymbol{\downarrow}$ 4.75, $\boldsymbol{\downarrow}$ 6.06\\
     \bottomrule
\end{tabular*}
\label{table:CH4 Trans}
\end{table*}

There are also more absorption features in the 1800 -- 1100 cm$^{-1}$ (5.50 -- 9.55 ${\mu}$m) range (Figure~\ref{fig:TotalTrans}, top panel). This is considered the fingerprint region of the spectrum, where a combination of organic molecules and various stretching and bending modes are present. The fingerprint region is highly complex, but it is also unique to each combination of molecules. This can better help determine the molecular structure of the haze sample pre-- and post--irradiation. There is a large drop in the transmittance of the sample between 1800 -- 1500 cm$^{-1}$ (5.50 -- 6.50 ${\mu}$m) due to a combination of C=C and C=O stretching (Figure~\ref{fig:TotalTrans}, top panel). Between 1126 -- 1100 cm$^{-1}$ (8.88 -- 9.06 ${\mu}$m), there are larger relative changes in the C--O stretch spectral feature during irradiation than in the rest of the spectrum (Figure~\ref{fig:TotalTrans}, top panel). During irradiation across the 350 nm spectrum, there is a factor of 0.12 difference between pre-- and post--irradiation. This is larger than in the 228 nm filter, where only a factor of 0.04 change is seen. For this specific spectral feature, the lower--energy simulated flare creates a larger increase in transmittance, meaning that less C--O stretching is present. We suggest that a potential mechanism for the decrease in C-O stretching may be the dissociation of the complex organics present in the haze sample.

\subsubsection{CO--Derived Haze Sample}
The 5\% CO atmosphere haze sample in transmittance, Table \ref{table:CO Trans} and Figure \ref{fig:TotalTrans}, bottom panel, identifies the bonds responsible for each spectral feature in Figure \ref{fig:TotalTrans}, bottom panel. There are characteristic bonds of O--H, C--H, C=O, C--O, and N--H. Overall, there are not large changes between pre--irradiation and post--irradiation across the 228 nm spectrum. There are larger changes across the 350 nm spectrum. The continuum has a lower transmittance longward of 2700 cm$^{-1}$ (3.6 $\mu$m), and then increases in transmittance until approximately 1350 cm$^{-1}$ (7.5 $\mu$m) (Figure \ref{fig:TotalTrans}, bottom panel). In addition, the total transmittance of the 5\% CO--derived haze is between 91\% to 100\%, which is a narrower range compared to the 5\% CH$_4$ atmosphere haze sample. This is likely due to the fact that the CO--derived haze sample is a thinner film compared to the CH$_4$--derived haze, in addition to any compositional differences between the two samples.  

One distinction between the two samples is that there is no large absorption feature corresponding to an O--H stretch in the 5\% CO--derived haze sample (Figure \ref{fig:TotalTrans}, bottom panel). One potential mechanism for this is the CH$_4$ present in the sample more readily dissociates (e.g., 360 kJ/mol or 0.33 $\mu$m; \citealt{cottrell1954strengths}), while CO bonds require higher energies (e.g., 1072kJ/mol or 0.111 $\mu$m; \citealt{cottrell1954strengths}) to be broken. The dissociated CH$_4$ creates excess hydrogen for other molecules to recombine with to form more complex O--H bonds. There are two different peaks at 2925 cm$^{-1}$ (3.42 $\mu$m) and 2850 cm$^{-1}$ (3.51 $\mu$m), both indicating the presence of C--H stretching due to alkanes present in the haze sample (Figure \ref{fig:TotalTrans}, bottom panel). After correcting the baseline for analysis, we find that during the irradiation process, the higher--energy filter slightly increases the transmittance of the spectral feature, while the lower--energy filter slightly decreases the spectral feature. This is similar to the 5\% CH$_4$ atmosphere haze sample, although the 5\% CH$_4$ atmosphere haze sample shows slightly larger relative changes during irradiation (Figure \ref{fig:TotalTrans}). 

\begin{table*}[ht]
\centering
\caption{Spectral features and corresponding functional groups identified in the 5\% CO--derived atmosphere haze sample 
pre-- and post--irradiation in transmittance and reflectance. Spectral features with an * are seen in both spectra.} 
\textbf{\large Transmittance}
\begin{tabular*}{\linewidth}{lllll}
\toprule
\multicolumn{1}{p{2cm}}{\centering \textbf{\large Frequency \\ (cm$^{-1}$)}}
& \multicolumn{1}{p{2.3cm}}{\centering \textbf{\large Wavelength \\ (${\mu}$m)}}
& \multicolumn{1}{p{3.8cm}}{\centering \textbf{\large Functional Group \\ (bond, type)}}
& \multicolumn{1}{p{2.5cm}}{\centering \textbf{\large Intensity}}
& \multicolumn{1}{p{3.7cm}}{\centering \textbf{\large Relative Change \\ (228 nm, 350 nm)}}
\\ \midrule
2925 -- 2850  & 3.42 -- 3.51  & *C--H Stretching (alkane) & medium, doublet & no changes\\
2175  & 4.63  & C$\equiv$N stretching (conjugated) & strong, broad & no changes \\
1620 -- 1580  & 6.2 -- 6.35  & N--H bending (amine) & medium, broad & no change, $\boldsymbol{\uparrow}$ 0.04\\
1390 -- 1320  &  7.25 -- 7.66  & O--H bending (alcohol) & weak, broad & no change, $\boldsymbol{\uparrow}$ 0.03\\
1150 -- 1100  &  8.72 -- 9.07  & C--O stretching (alcohol) & strong, broad & no change, $\boldsymbol{\uparrow}$ 0.04\\

\midrule
\\ \multicolumn{5}{c}{\textbf{\large Reflectance}} \\
\midrule

3125 -- 3075  & 3.2 -- 3.25  & O--H alcohol (intramolecular bonded) & weak, broad & no changes\\
2925 -- 2850  & 3.42 -- 3.51  & *C--H stretching (alkane, alkene) & medium, shoulder & no change, $\boldsymbol{\downarrow}$ 0.03\\
2260 -- 2222  & 4.42 -- 4.5  & C$\equiv$N stretching (nitrile group) & weak, broad & no changes\\
1875 -- 1845  & 5.32 -- 5.42  & C=O stretching (anhydride) & medium, sharp & no change, $\boldsymbol{\downarrow}$ 0.02\\
1740 -- 1700  & 5.75 -- 5.87  & C=O stretching (aldehyde) & strong, broad & no change, $\boldsymbol{\downarrow}$ 0.04\\
1460 -- 1415  & 6.85 -- 7.06  & C--H bending (alkane, methyl group) & medium, broad & no change, $\boldsymbol{\downarrow}$ 0.04\\
1275 -- 1220  & 7.85 -- 8.17  & *C--O stretch (ether group) & strong, broad & $\boldsymbol{\uparrow}$ 0.03, $\boldsymbol{\downarrow}$ 0.07\\
     \bottomrule
     
\end{tabular*}
\label{table:CO Trans}
\end{table*}

Another difference between the two sample spectra is that there is no large drop in transmittance between 1800 -- 1500 cm$^{-1}$ (5.50 -- 6.50 $\mu$m) (Figure \ref{fig:TotalTrans}, bottom panel). This may be due to a lack of dissociation between the C$\equiv$O bonds, which need significantly more energy to dissociate as compared to the C--H bonds found in CH$_4$ atmosphere haze. Instead, the largest drop in the 5\% CO atmospheric haze sample is between 1390 -- 1320 cm$^{-1}$ (7.25 -- 7.66 $\mu$m), where O--H bending is present (Figure \ref{fig:TotalTrans}, bottom panel). Across this spectral feature, the 350 nm filter increases with wavelength, whereas it decreases across the 228 nm filter. Both changes can be attributed to the overall continuum changes rather than any specific spectral features, indicating a general change of the haze overall rather than any specific functional group. 

\subsection{Reflectance Spectroscopy Results}\label{Refresults}
\begin{figure}[h!] 
    \centering
    \includegraphics[angle=-90, width=\columnwidth]{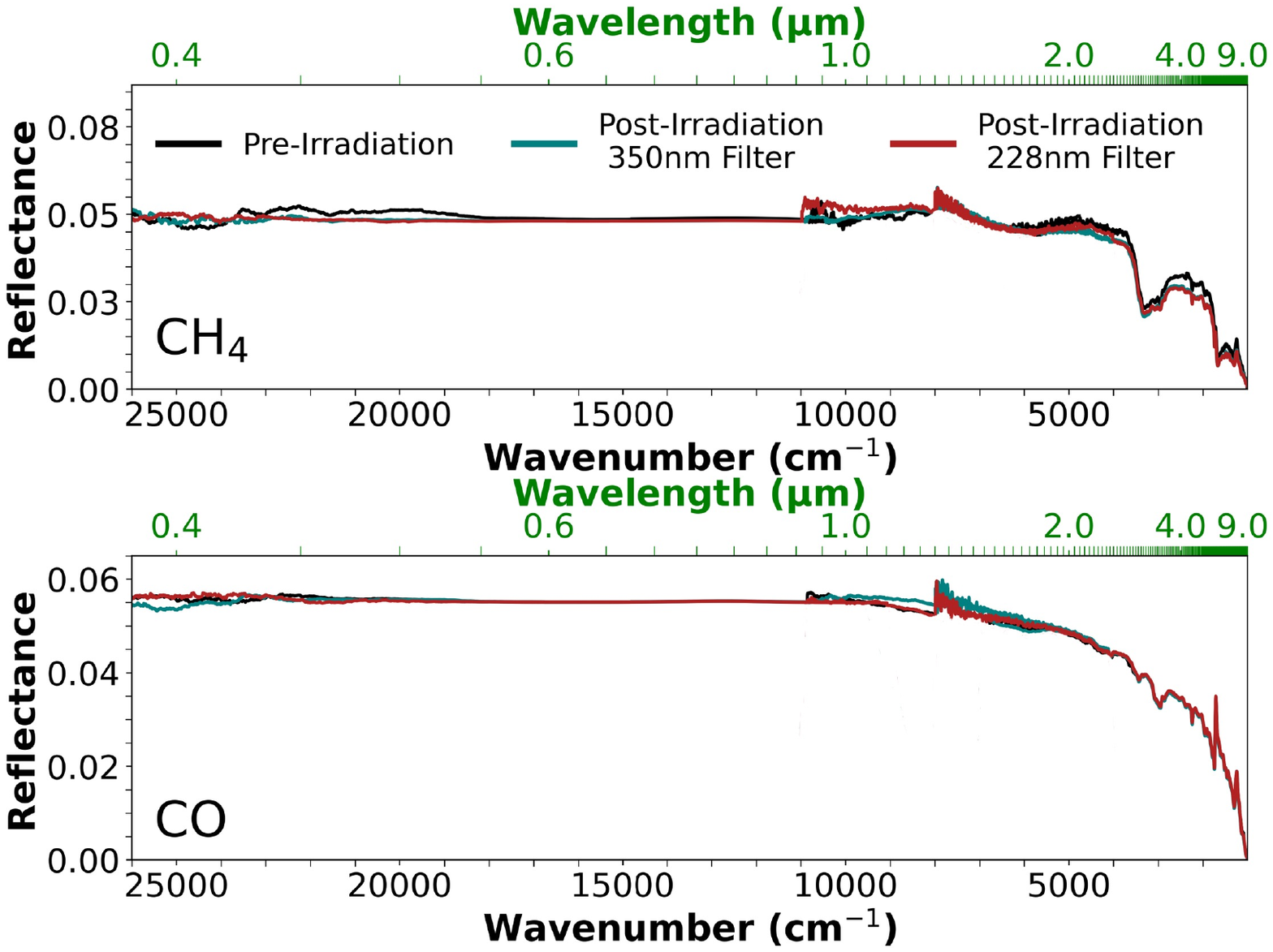}
    \caption{Reflectance spectra of our 5\% CH$_4$--derived haze sample (top) and 5\% CO--derived haze sample (bottom) in the visible to mid--IR wavelength region (26000 -- 1000 cm$^{-1}$, 0.4 -- 9 $\mu$m). The black line represents the samples before the irradiation process began, and the teal and red lines represent the post--irradiation spectrum across the 350 nm and 228 nm filters respectively.}
    \label{fig:Reflectance_total}
\end{figure}

\begin{figure}[h!] 
    \centering
    \includegraphics[angle=-90, width=\columnwidth]{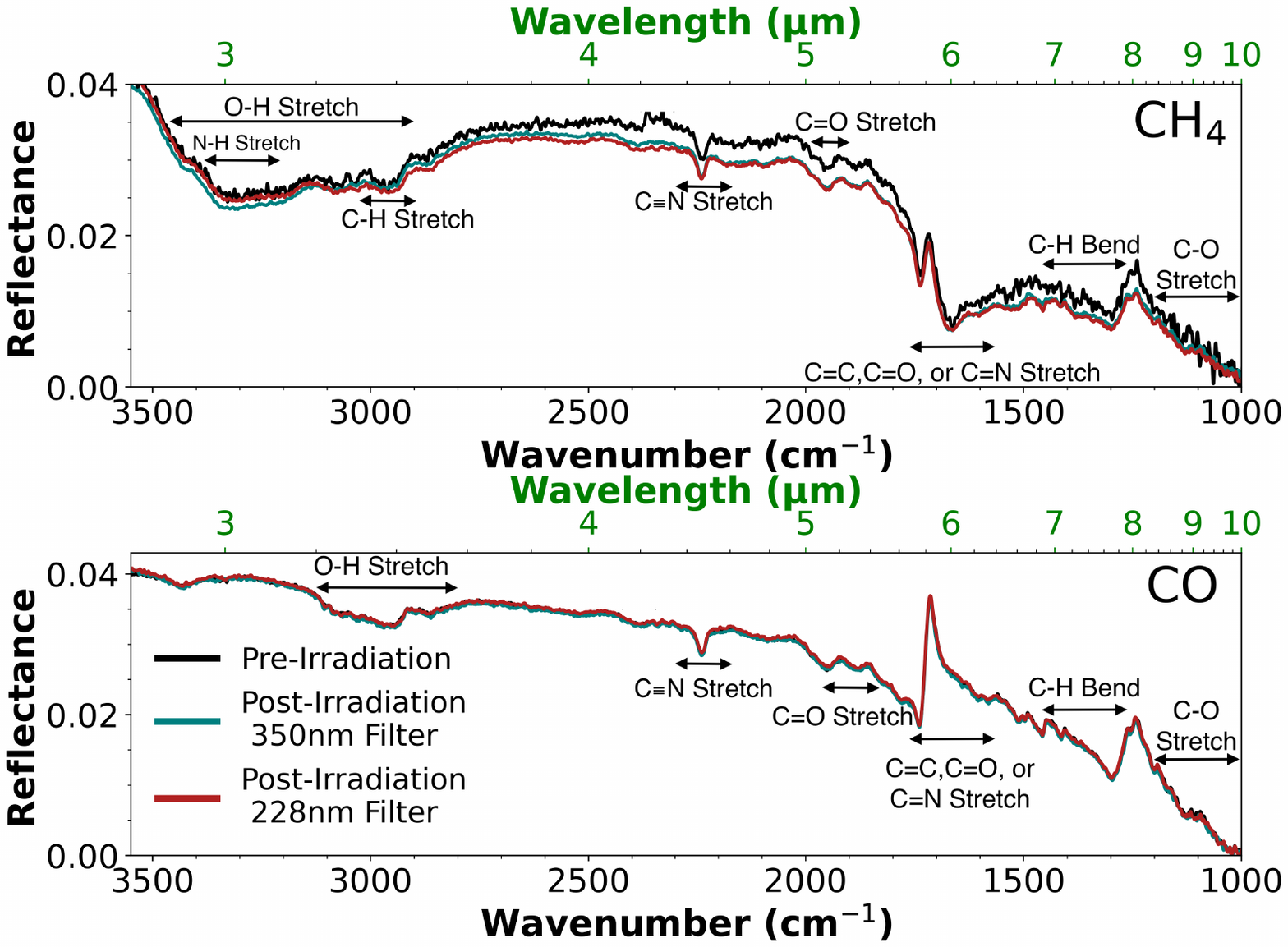}
    \caption{Enlarged spectrum between 3500 -- 1000 cm$^{-1}$ (2.5 -- 10 $\mu$m) of the 5\% CH$_4$ (top) and 5\% CO (bottom) atmosphere generated hazes as a function of wavelength in reflectance with major spectral features labeled. Between pre-- and post--irradiation, there are larger changes in the 5\% CH$_4$ derived haze spectrum in comparison to the 5\% CO derived haze spectrum, where the differences are not obvious.}
    \label{fig:TotalRef}
\end{figure}

Figure \ref{fig:Reflectance_total} shows the reflectance spectra of both samples pre-- and post--irradiation. The general spectral shape and features are similar throughout the irradiation process. We note that the thickness of the haze films, material properties, and the chemical compositions of the films impact the reflectance measurements in this experiment (e.g., \citealt{He2023Methods}). Thicker haze films will generally, but not always, be more reflective in the overall spectral continuum. Specific functional groups will also generally change the reflectance depending on the abundance of the species at that wavelength. (e.g., \citealt{He2018HazeRough}). The spectrum across both samples in Figure \ref{fig:Reflectance_total} are relatively flat and featureless from 0.4 to 2.5 ${\mu}$m. The continuum stays flat with no increasing or decreasing slopes. More spectral features appear between 2.5 to 9 ${\mu}$m due to various organic functional groups found in the exoplanet haze samples (Figure \ref{fig:TotalRef}). The results are broken down by sample. 

\begin{figure}[h] 
    \centering
    \includegraphics[width=\columnwidth]{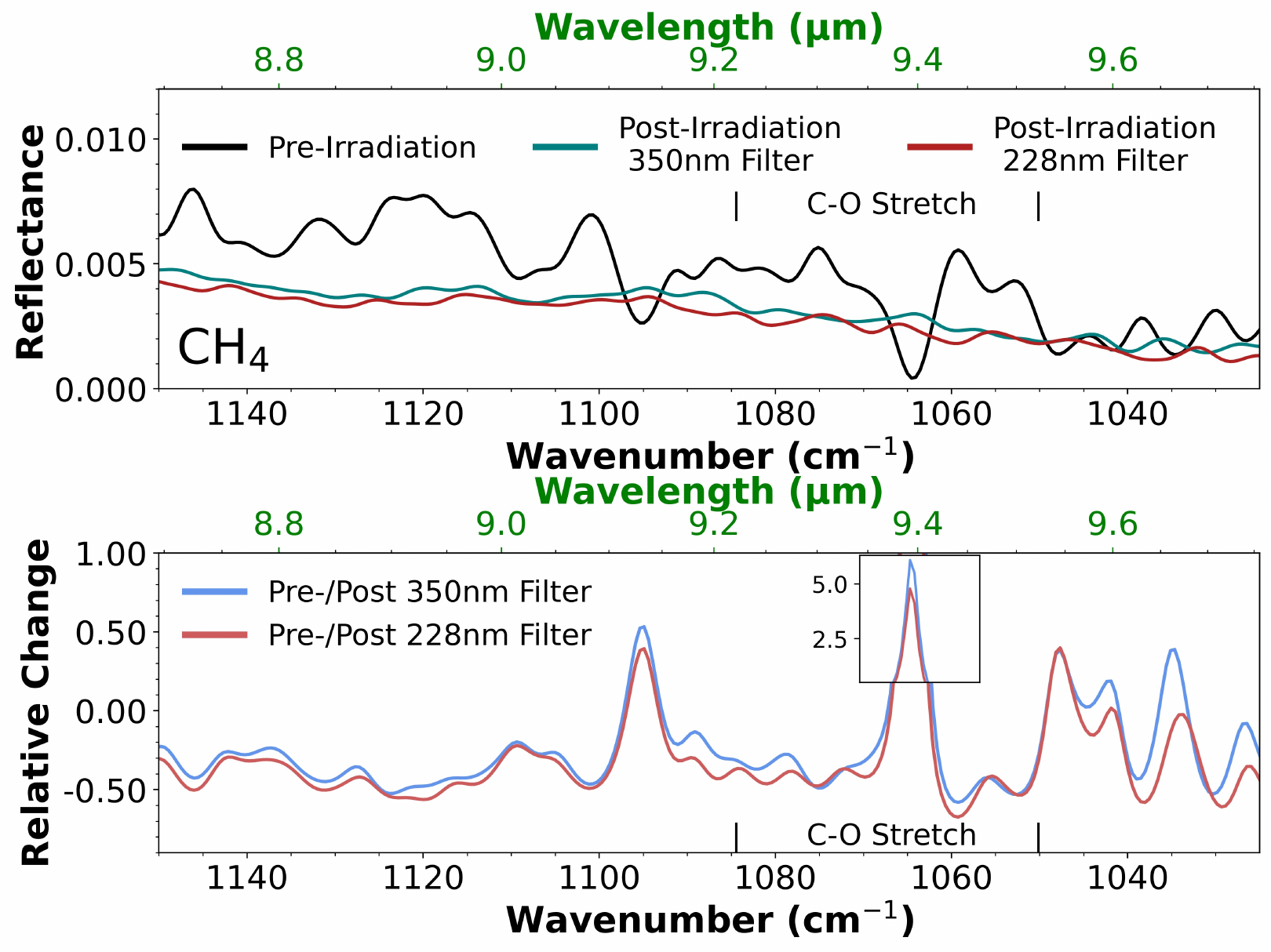}
    \caption{Top: The C--O stretching features of the 5\% CH$_4$ derived haze sample evident of a primary alcohol seen pre-- and post--irradiation in reflectance. The irradiation process made the spectral features less pronounced in relation to the pre--irradiation spectra across both filters. Bottom: the relative change between pre--irradiation and post-irradiation of each filter respectively.}
    \label{fig:1050-85Feature}
\end{figure}

\subsubsection{CH$_4$--Derived Haze Sample}
In the 5\% CH$_4$ atmosphere haze sample, Table \ref{table:CH4 Trans} identifies the bonds responsible for each spectral feature in Figure \ref{fig:TotalRef}, top panel. There are characteristic features of O--H, C--H, C=O, C--O, C=N, and C=C bonds. The large absorption feature between 3400 -- 2700 cm$^{-1}$ is due to O--H stretching, which is indicative of both alcohols (3500 -- 3200 cm$^{-1}$) and carbonic acids (3300 -- 2700 cm$^{-1}$) in this haze sample (Figure \ref{fig:TotalRef}, top panel). Within this feature there are also both N--H (3400 -- 3225 cm$^{-1}$) and C--H (3075 -- 2700 cm$^{-1}$) bond stretching. The spectral features decrease during the irradiation process. There are also more absorption features between 1740 -- 1050 cm$^{-1}$ in the fingerprint region (Figure \ref{fig:TotalRef}, Top panel). During the irradiation process, both filters decrease these spectral features across the spectrum. Of particular interest is the C--O bond stretching between 1085 -- 1050 cm$^{-1}$ seen in Figure \ref{fig:1050-85Feature}, top panel. There are pronounced changes between pre-- and post--irradiation, and the spectrum across both filters becomes flatter than pre--irradiation in comparison for a factor of 4.75 change in the 228 nm filter and a factor of 6.06 change in the 350 nm filter (Figure \ref{fig:1050-85Feature}, bottom panel). We note that the pre--irradiation value for this feature is anomalously low, leading to a large relative change in the C--O spectral feature. However, this indicates a potential for compositional changes in the haze sample, and a potential mechanism for loss or dissociation of water--world hazes during stellar flaring events.

\subsubsection{CO--Derived Haze Sample}
 In the 5\% CO atmosphere haze sample, Table \ref{table:CO Trans} identifies the bonds responsible for each spectral feature in Figure \ref{fig:TotalRef}, bottom panel. There are characteristic O--H, C--H, C=O, C--O, C$\equiv$N, and C=C bonds, as with the transmittance measurements and seen in the 5\% CH$_4$ derived haze sample. The overall reflectance of the sample is between 0\% to 4.5\%. The absorption feature between 3500 -- 3000 cm$^{-1}$ is largely due to O--H bond stretching, with a small contribution from a N--H amine stretch (Figure \ref{fig:TotalRef}, bottom panel). This is a weak, broad feature that indicates intramolecular bonded alcohols. The spectral features decrease during the irradiation process. The stretching of C--H between 2925 -- 2850 cm$^{-1}$ is present as a medium doublet in the spectrum (Figure \ref{fig:TotalRef}, bottom panel). The two large peaks in the sample between 1740 -- 1700 cm$^{-1}$ and 1275 -- 1220 cm$^{-1}$ are C=O and C--O stretches, which indicate the presence of complex organics (Figure \ref{fig:TotalRef}, bottom panel) as seen in the 5\% CH$_4$ derived haze sample. The C=O stretch feature remains unchanged during the irradiation process. The C--O stretch decreases under the 350 nm filter. Other absorption features in the spectrum indicate the presence of nitriles, aromatics, and organic groups. Largely the 228 nm filter increases the spectral feature across the haze sample, and the 350 nm filter decreases the spectral feature across the haze sample (Figure \ref{fig:TotalRef}, bottom panel). The differences during irradiation between the two filters lead to distinctive compositional changes. We hypothesize that this compositional change is driven by the different energies allowed by each filter, as chemical reactions unique to the haze samples will occur, leading to distinctive compositions after UV bombardment. These spectral and compositional changes will alter the resulting effectiveness of the haze in insulating the planetary atmosphere from the high--radiation events of stellar flaring. 
\begin{figure*}[ht!]
    \centering
    \includegraphics[width=0.8\textwidth]{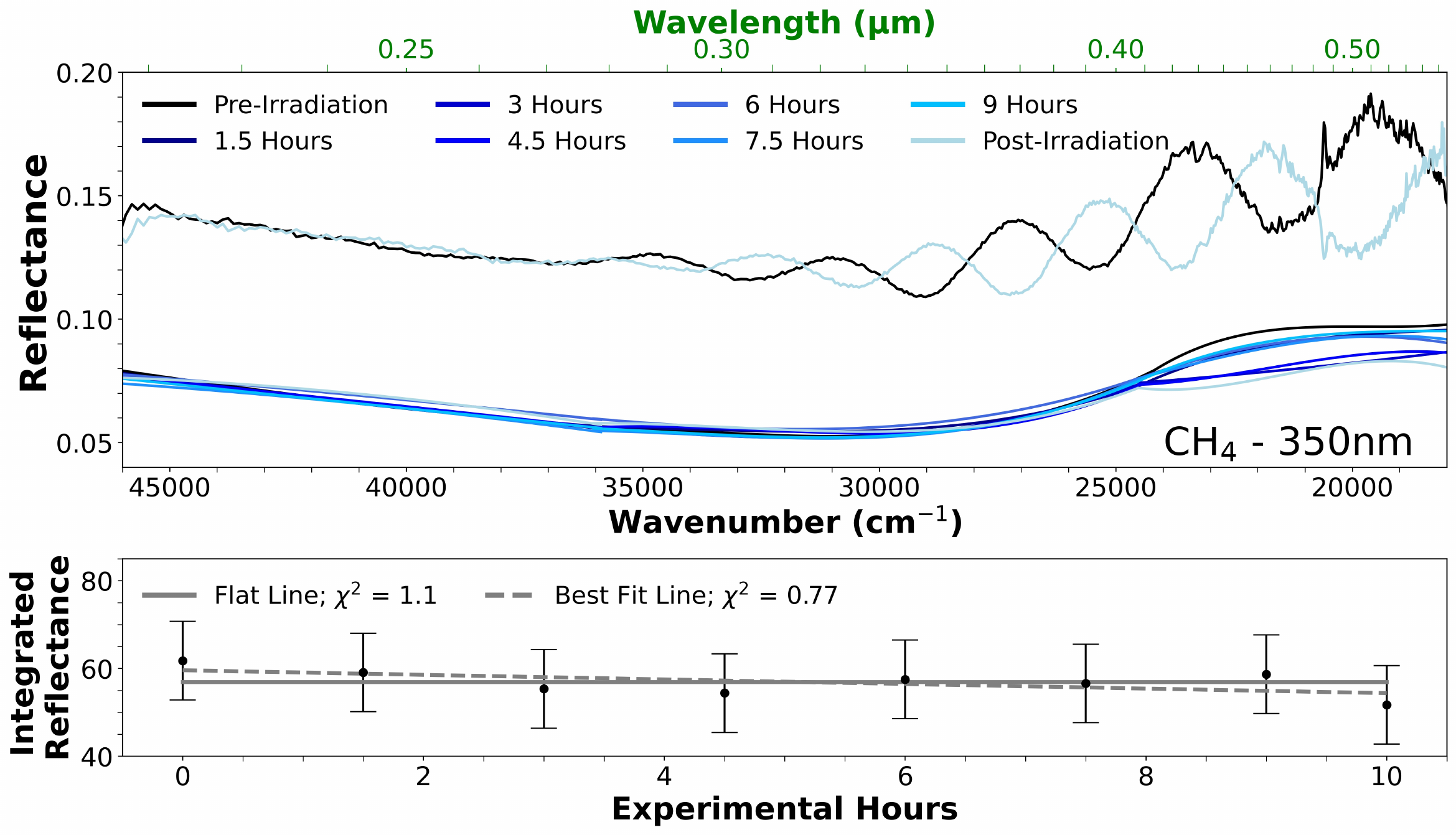}
    \caption{Top: UV-visible (46000 -- 18000 cm$^{-1}$, 0.22 -- 0.54 ${\mu}$m) spectrum for the 5\% CH$_4$ atmosphere haze sample across the 350 nm filter, corrected for interference fringe effects. The pre-- and post--irradiation uncorrected spectra have been offset vertically by 0.065 for clarity. The corrected and smoothed data during the irradiation process are plotted as is. Bottom: The total integrated reflectance of the fringe--corrected smoothed data spanning the plotted wavelength range over the duration of the irradiation process. The gray solid line is the best fit flat line, with a reduced $\chi^2$ of 1.1. The gray dashed line is the best fit sloped line, with a reduced $\chi^2$ of 0.77. The error bars for the integrated reflectance are derived from propagating error throughout the instrument, fringe interference equation, smoothing, and integration.}
    \label{fig:CH4_320}
\end{figure*}
\subsection{Transmittance and Reflectance Summary}\label{TRsummary}
As shown in Tables \ref{table:CH4 Trans} and \ref{table:CO Trans} and Figures \ref{fig:TotalTrans} and \ref{fig:TotalRef}, there are more spectral features observed in the reflectance spectra of both samples in comparison to transmittance. There are also more spectral features observed in the 5\% CH$_4$--derived haze sample than the 5\% CO--derived sample. This means that the laboratory haze samples are compositionally different. 

In both transmittance and reflectance, stretches of O--H, C--H, C--O, and C$\equiv$N bonds are present in the 5\% CH$_4$--derived haze sample (Figures \ref{fig:TotalTrans}, \ref{fig:TotalRef}, top panel). There are more distinguishable features in the reflectance spectrum as compared to transmittance, which may be due to the fact that reflectance spectroscopy enhances weak features that may not be seen in transmission spectroscopy. We also observe C--O, C=O, and C--H stretching in the reflectance spectrum of the CH$_4$ derived haze sample (Figure \ref{fig:TotalRef}, top panel), as many of these stretches may be present in transmission, but could be overshadowed by larger features. 

Large spectral features of C--H and C--O bonds are present in the 5\% CO--derived haze sample in transmittance (Figure \ref{fig:TotalTrans}, bottom panel). In this case, the number of spectral features observed in  reflectance is only slightly higher than in transmittance. However, there are some features that do not match between spectra, which may be due to the stretching bands being shifted to higher wavenumbers in reflectance \footnote{\href{https://www.photonics.com/images/Web/WhitePapers/532/Andor_Learning_Absorption_Transmission_Reflection_Spectroscopy_For_Photonics.pdf}{Absorption / Transmission / Reflection Spectroscopy}}. 

 This means that the laboratory haze samples are compositionally different not only between the two different derived haze samples, but also pre-- and post--irradiation. This has key implications in the retainability of water--world exoplanet hazes and how UV energy can change the compositions and overall reflectivity of the hazes over short timescales. 
\begin{figure*}[ht]
    \centering
    \includegraphics[width=0.8\textwidth]{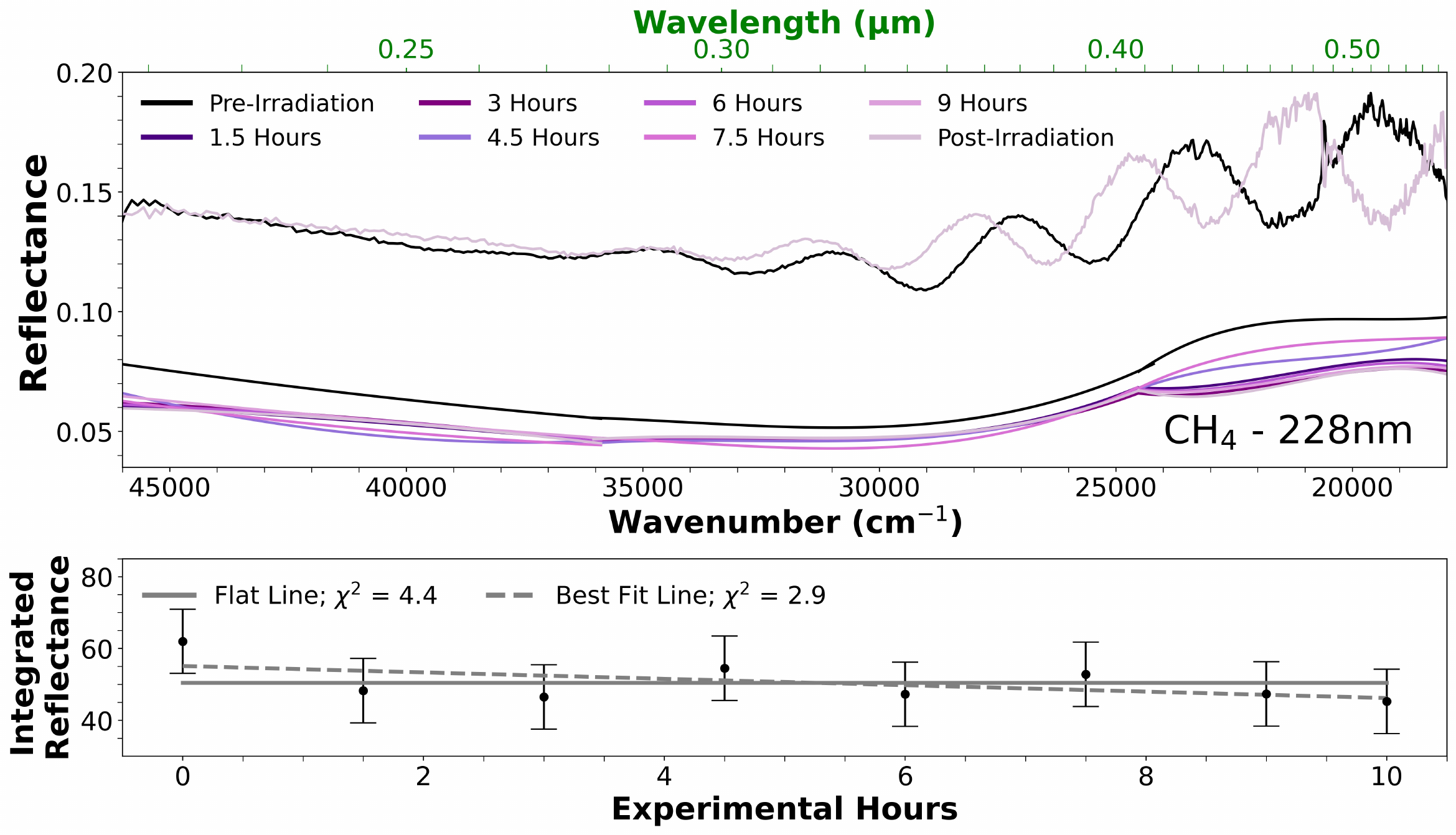}
    \caption{Top: UV-visible (46000 -- 18000 cm$^{-1}$, 0.22 -- 0.54 ${\mu}$m) spectrum for the 5\% CH$_4$ atmosphere haze sample across the 228 nm filter. The pre-- and post--irradiation uncorrected spectra have been offset vertically by 0.065 for clarity. The corrected and smoothed data during the irradiation process are plotted as is. Bottom: The total integrated reflectance of the fringe--corrected and smoothed data spanning the plotted wavelength range over the duration of the irradiation process. The gray solid line is the best fit flat line, with a reduced $\chi^2$ of 4.4. The gray dashed line is the best fit sloped line, with a reduced $\chi^2$ of 2.9. The error bars for the integrated reflectance are derived from propagating error throughout the instrument, fringe interference equation, smoothing, and integration.}
    \label{fig:CH4_240}
\end{figure*}

\subsection{UV-Vis Time Dependent Results}\label{TDresults}
\subsubsection{CH$_4$--Derived Haze Sample}
In addition to pre-- and post--irradiation spectra, we obtained UV-visible reflectance data every 1.5 hours throughout the 10 hour bombardment for the CH$_4$--derived haze sample (Figures \ref{fig:CH4_320} and \ref{fig:CH4_240}). We did not obtain transmittance data in this wavelength range, as we would have needed to break N$_2$ purge. The 5\% CH$_4$--derived haze sample began as a thicker film based on the thickness calculations (Table \ref{table:Thickness}) and interference fringes present both pre-- and post--irradiation. We also hypothesize the sample was thicker from the post--irradiation magnified images. However, we do not take the composite images into account when calculating the sample thickness. We show the spectra changes over time in both Figure \ref{fig:CH4_320}, which corresponds to the 350 nm filter, and Figure \ref{fig:CH4_240}, which corresponds to the 228 nm filter. After the irradiation process, the interference fringes shift to shorter wavelengths in both filters. The actual thickness of the haze film changes post--irradiation, from 0.911 $\pm$ .094 $\mu$m pre--irradiation to 0.862 $\pm$ .021 $\mu$m across the 350 nm filter, and 0.891 $\pm$ .056 $\mu$m across the 228 nm filter, meaning the irradiation process physically alters the haze layers present on the MgF$_2$ plate (Table \ref{table:Thickness}). This is also apparent with slight continuum changes in the infrared region both in reflectance and transmittance (e.g., Figures \ref{fig:TotalTrans}, \ref{fig:TotalRef}). 

Figure \ref{fig:CH4_320} shows the time--series data throughout the irradiation process for the CH$_4$--derived haze sample across the 350 nm filter. After correcting for the optical fringes as described in Equation (\ref{eq:Interference equation}), the corrected spectra were plotted as a function of wavelength. Due to the large amplitude changes towards longer wavelengths, the corrected data becomes slightly noisier as wavelength increases. We fit a polynomial smoothed line to the data to eliminate the noise for our total reflectance integration. The spectrum is flat throughout the irradiation process, and no specific spectral features can be discerned. To see any potential changes in the overall reflectance of the sample, we integrate our smoothed spectrum across the plotted wavelength range as a function of experimental hours (Figure \ref{fig:CH4_320}, bottom panel). We then fit a flat line and sloped line to the integrated data, and perform both a reduced $\chi^2$ and Bayesian Information Criterion (BIC) analysis. We find that our best fit flat line had a reduced $\chi^2$ value of 1.1, and the best fit sloped line had a reduced $\chi^2$ value of 0.77, in which our sloped line slightly fits the data best, though it is overfit. If we carry out the BIC test comparing the two line fits, we get a $\Delta$BIC value of 3.6, which would slightly prefer the sloped line fit indicating a decrease in reflectance. However, because the reduced $\chi^2$ value of the best fit sloped line is less than 1, we cannot conclude that one is preferred. We change the best fit sloped line to a reduced $\chi^2$ value of 1 to remove the over fitting for further analysis, and get a $\Delta$BIC value of 1.75. Although we obtained a better reduced $\chi^2$ for the sloped line indicating a slight degradation in the total integrated reflectance over time, this is not statistically preferred over the no change line within the error bars. 

\begin{figure*}[ht]
    \centering
    \includegraphics[width=0.8\textwidth]{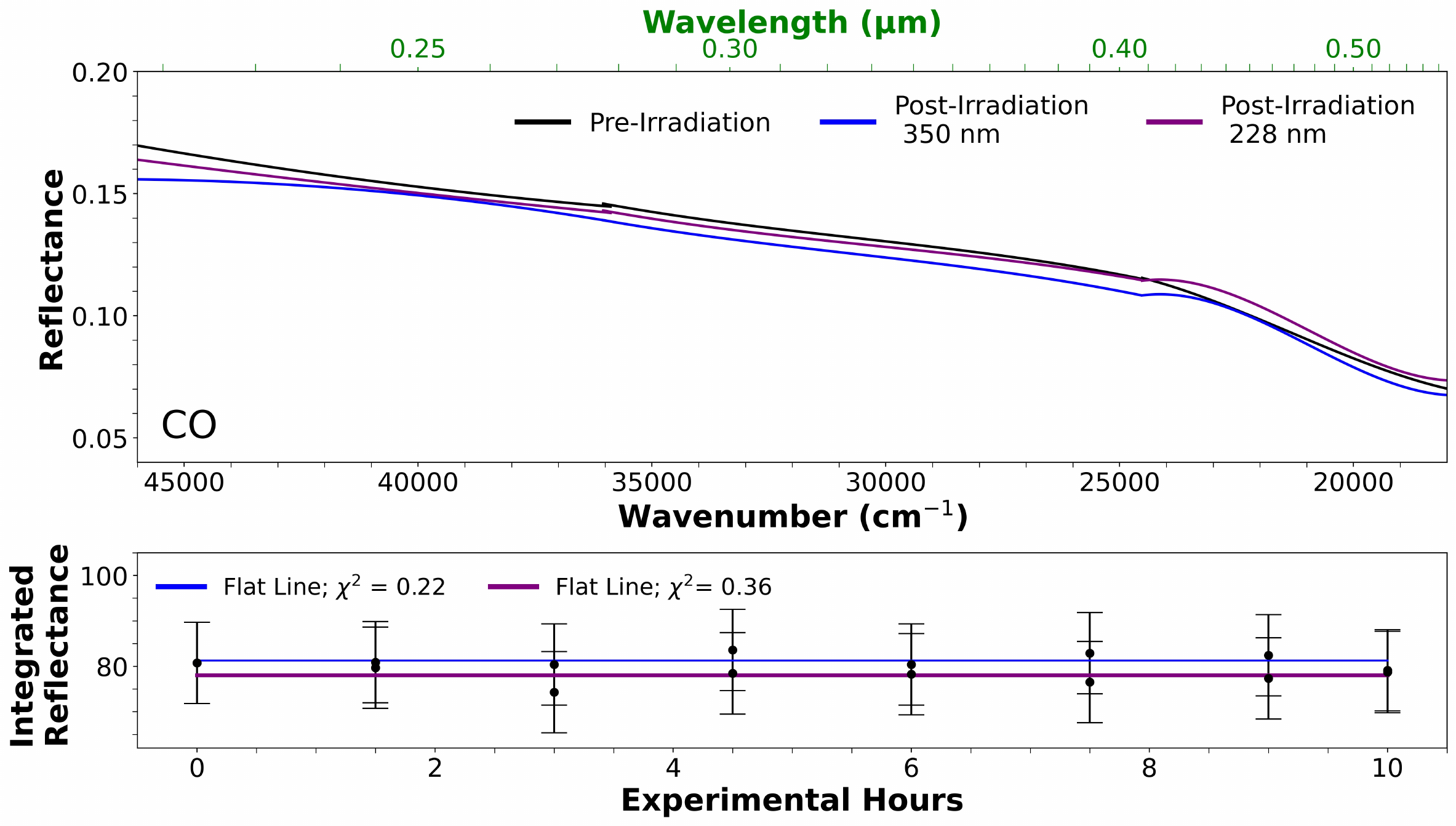}
    \caption{Top: UV-visible (46000 -- 18000 cm$^{-1}$, 0.22 -- 0.54 ${\mu}$m) smoothed spectrum for the 5\% CO atmosphere haze sample across both the 228 nm and 350 nm filters pre-- and post--irradiation. Due to the apparent thinness of the haze film, there is not enough fringing seen to be corrected. Bottom: The total integrated reflectance of the smoothed data spanning the plotted wavelength range over the duration of the irradiation process. The blue solid line is the best fit flat line for the 350 nm filter, with a reduced $\chi^2$ of 0.22. The purple dashed line is the best fit flat line for the 228 nm filter, with a reduced $\chi^2$ of 0.36. The error bars for the integrated reflectance are derived from propagating error throughout the instrument, fringe interference equation, smoothing, and integration. There are no changes within the error bars in the integrated film reflectance between pre-- and post--irradiation in this wavelength range.}
    \label{fig:CO-UVTotal}
\end{figure*}

Figure \ref{fig:CH4_240} shows the time--series data throughout the irradiation process for the CH$_4$--derived haze sample across the 228 nm filter. After correcting for the optical fringes, the corrected spectrum was smoothed and plotted as a function of wavelength. We then integrate our smoothed spectra across the plotted wavelength range as a function of experimental hours (Figure \ref{fig:CH4_240}, bottom panel). Like the 350 nm filter, we fit a flat line and sloped line to the integrated data, and perform both a reduced $\chi^2$ and BIC analysis. We find that our best fit flat line had a reduced $\chi^2$ value of 4.4, and the best fit sloped line had a reduced $\chi^2$ value of 2.9, in which our sloped line fits the data best. We further verify our results by carrying out the BIC text comparing the two line fits, and get a $\Delta$BIC value of 3.1, which prefers the sloped line fit. Therefore we conclude that over the experimental time, we see a slight decrease in the total integrated reflectance. This change in integrated reflectance suggests that our sample across the 228 nm filter became less reflective over the irradiation process, leading to an increase in transmittance. This is more significant of a decrease than the 350 nm filter, where we do not see a statistically significant change in reflectance. This conclusion contrasts the change in film thickness, as seen in Table \ref{table:Thickness}, where the haze after the 350 nm filter is slightly thinner than after the 228 nm filter. However, we note that the pre-- and post--thicknesses are within error bars of each other, so this apparent contrast between change in reflectance and thickness as a function of filter wavelength are not strongly in tension. Some potential hypotheses for this film alteration in reflectance across both filters is the dissociation of the haze molecules in addition to a change in the material properties. 

\subsubsection{CO--Derived Haze Sample}
Figure \ref{fig:CO-UVTotal} shows the pre-- and post--irradiation spectra of the 5\% CO atmosphere haze sample for the ultraviolet--to--visible region. In the UV--visible wavelength region, there are no changes in reflectance or in the interference fringing during the irradiation process for the 5\% CO--derived haze sample. This is different from the 5\% CH$_4$--derived haze sample, where we do see a decrease in reflectance pre-- and post--irradiation across the 228 nm filter. The CO-derived haze sample was thinner before beginning our experiments based upon the haze thickness calculations (Table \ref{table:Thickness}) and also suggested by the post--irradiation magnified images (Figure \ref{fig:HazePictures}), and only slight interference fringes are seen. Due to the limited amount of fringes and lack of large amplitudes, they were not corrected for in this analysis. In addition, the potential interference fringes do not change or shift during the irradiation process (Figure \ref{fig:CO-UVTotal}, top panel). A calculated film thickness of 0.368 $\pm$ .037 $\mu$m both pre-- and post--irradiation shows that the film thickness did not change over the irradiation process. In addition, after performing a reduced $\chi^2$ analysis, we find that both lines have reduced $\chi^2$ values less than 1, 0.22 for the 350 nm filter and 0.36 for the 228 nm filter respectively, meaning that we are over--fitting and that there is no change seen across the experimental timeline for either filter. Some potential hypotheses for this lack of film thickness and reflectance change is that either the film was too thin initially to react to the UV bombardment to see any observed change, or that the material properties of the sample changed to compensate for any change in reflectance during the irradiation process, though the later scenario is unlikely (Figure \ref{fig:CO-UVTotal}, bottom panel). 

We conclude that the simulated flare in this work did not change the thickness or total reflectance of the 5\% CO haze sample, and slightly changed the total reflectance of the 5\% CH$_4$ sample over the irradiation process in the 228 nm filter, with the 350 nm filter degradation not being statistically significant. If we apply this to a planetary analogue, we can conclude that there is a possibility that high energy stellar activity would slightly decrease the amount of haze present in a water--world atmosphere via destruction. The current picture in the prebiotic chemistry community is that organics irradiated using UV radiation become “tar” (e.g., \citealt{BennerPrebiotic}). This contrasts with the standard picture in the Mars biosignatures community which is that irradiated organics become gas (e.g., \citealt{Dartnell2014MarsUV,Carrier2019MarsUV}, which our work favors more.

This suggests that stellar flaring has the potential to have a large effect on atmospheric hazes, and could strip away haze from the atmosphere of a water--world exoplanet. We note that this experiment does not take into account any haze generation completed via UV irradiation, only destruction of haze already made. In a real atmosphere, the UV radiation might change the physical and chemical properties of the haze particles, but the particles could still be present in the atmosphere.

\section{\textbf{Discussion}}\label{sec:Discussion}

So far, haze interactions with UV light representing solar flares have been studied for Titan--like hazes and tholins (e.g., \citealt{Gavilan2018Haze,Carrasco2018TitanUV}). 
Some of the spectral features (e.g., CO$_2$, CO, CH$_4$) seen in this work are documented throughout the literature {(e.g., \citealt{Tran2008TitanHaze,Horst2014HazeCO,Horst2017Titan,Fleury2017CO2Atmo,Gavilan2018Haze,Fleury2019H2Atmo,Jovanovic2020PlutoAerosol,He2022Titan,Moran2022Triton,He2023Methods}). However, these concepts of haze--irradiation interaction and haze spectral features have not yet been combined to study the changes in spectral features due to UV irradiation on laboratory--made exoplanet hazes.

This work provides a comprehensive spectral analysis of laboratory--made exoplanet hazes exposed to UV light over a large wavelength range. To examine how representative our experiments are to an actual planetary atmosphere experiencing flares, we compare our experimental setup to a typical flare of an M dwarf star. The power output of the UV light is 1.1 W m$^{-2}$ for a 10 hour time period. The quiescent period radiation of an M dwarf star (e.g., GJ~1214) is approximately 3.5 W m$^{-2}$ \cite[]{He2023Methods}. This is three times stronger radiation output than our simulated flare. In addition, flaring energies and photoelectrons (e.g., \citealt{Pirim2015photoelectrons}) have significantly more radiation than quiescent energies, including high--energy UV and X--Ray radiation. The total energy produced by M dwarf flares can be orders of magnitude higher than quiescent energies (e.g., \citealt{Loyd2016Flare, Howard2020FlareHab}). Therefore, we expect larger changes in both the continuum and specific spectral features than found in Section \ref{Results} in higher--energy environments. This study provides a baseline for simulated flaring effects on these planet hazes. Future work should include higher--energy lamps if feasible to fully understand the impact that high--energy UV output would have on exoplanet haze degradation. 

Our experiments focus on flare energy in the ultraviolet part of the electromagnetic spectrum. However, stellar activity, including flares, can release energy across a broad wavelength range, which can have even more effects on the orbiting planet's atmosphere and haze layer. Stellar activity temporarily increases the incident flux received by a planet, triggering not only photochemistry but atmospheric escape in the upper layers of the planet's atmosphere. More specifically, the X--ray/extreme ultraviolet (XUV, 100 -- 1000\AA) part of the stellar spectrum has the ability to penetrate into the haze layer if the hazes are found in the upper atmospheric layers. This work does not take into account XUV radiation, which may have detrimental results on the longevity of exoplanet hazes. Dissociation in the XUV wavelength region, including H$_2$, N$_2$, H$_2$O, CO, CO$_2$, and CH$_4$, occurs shortward of 117 nm \cite[]{Loyd2016Flare}. All of these molecules are found in the starting composition of our haze samples. By not including XUV radiation in our laboratory experiments, there is potential for further degradation and dissociation of hazes than the results shown here. In addition, this could also conversely provide more ions for further haze ion recombination mechanisms. The photodissociation rates and overall XUV flux in this range will vary between stars, and work to understand the stellar XUV region (e.g., \citealt{Linsky2014FlareXUV,Fontenla2016FlareEUV,Peacock2019,Duvvuri2021XUV}) in addition to the effects of the XUV region  (e.g., \citealt{Howard2020FlareHab,Foster2022Evaporation,Louca2023StellarActivity}) is already underway and may soon determine how XUV radiation contributes to the stripping of water--world hazes. 

Haze production and cycling in an exoplanet atmosphere has the ability to provide a shielding mechanism from stellar flaring \cite[]{Horst2018PHAZER}. This study does not account for the potential of haze production under UV light or via any other haze production mechanism. In this experiment, we assume that the representative photochemical haze is present and not continuing to grow during the simulated flaring event. Furthermore, previous laboratory experiments focusing on haze production rates show that they vary widely by atmospheric composition (e.g., \citealt{Horst2018PHAZER,He2018HazeRough,He2018MiniNeptune,Moran2022Triton}). This can have varying effects on the ability of the haze layer to be sustained over different timescales. Although we do not estimate the haze production rates, leaving this to a companion paper, the sample including CH$_4$ atmospheric haze visually produced more haze during the initial haze production process in comparison to the CO atmospheric haze. 

In addition, previous laboratory studies have shown that H$_2$O--dominated atmospheres produce more haze particles than either H$_2$-- or CO$_2$--dominated atmospheres \cite[]{Horst2014HazeCO,Horst2018PHAZER}. Therefore, we expect our two sample atmospheres to have higher production rates than other atmospheric hazes. However, the specific physical mechanisms which grow and produce organic haze particles in an exoplanet atmosphere are still largely unknown (e.g., \citealt{Horst2018PHAZER,He2020PHAZER,Moran2020SuperEarth,Vuitton2021}). There is potential for a high production rate to reach equilibrium with the haze degradation seen in this study. Alternatively, the radiation provided by the star could strip the haze particles away. It also remains unclear how a haze cycling mechanism could be sustained over evolutionary timescales. 


The hazes that will be observed in the future are expected to be of different atmospheric compositions and chemical makeups from each other, which will have different optical properties. Optical properties are necessary inputs for atmospheric modeling to better understand the associated physical and chemical processes at work. This work shows that irradiation of laboratory--made exoplanet hazes produces a different spectrum pre-- and post--irradiation, which will yield different haze optical properties. This work, paired with future optical constant calculations of these irradiated hazes, will be applicable for temperate water--world exoplanet atmospheres that orbit closely around M dwarf stars.

\section{\textbf{Conclusion}}\label{conclusion}
In this study, we subjected two water--world exoplanet haze samples to UV radiation through two different bandpasses of light to determine molecular changes and assess potential haze destruction. Our results show that irradiation alters both the overall continuum and specific spectral features of both the 5\% CH$_4$ and 5\% CO atmosphere derived samples. More specifically, the 5\% CH$_4$ atmosphere haze sample exhibits larger overall relative changes throughout each bandpass in both transmittance and reflectance. During the irradiation process in our time--dependent UV--visible wavelengths, we are able to quantify that across the 350 nm filter for the CH$_4$ sample, we did not find any statistically significant degradation. However, we did find a statistically significant decrease in the integrated reflectance across the 228 nm filter, leading to a less reflective haze layer on our representative disk. However, in comparing the haze thickness pre-- and post--irradiation, we note that the changes we observe for each filter are within 1$\sigma$ uncertainty across the 228 nm filter, and within 2$\sigma$ uncertainty over the 350 nm filter. Thus we cannot conclude that the changes in thickness are significant. In addition, this may also not be a direct correlation, as the reflectance of the sample is also affected by the surface roughness of the sample, which was not measured pre--irradiation, and the haze material properties. We also were able to quantify for the CO derived haze sample that there was no change during the irradiation process in integrated reflectance or haze thickness within the error bars.

This first study exposing exoplanet hazes to UV radiation provides a better understanding of continuum and spectral feature changes during the irradiation process over UV wavelengths. This study provides an avenue in which to determine and quantify spectral changes and resulting compositional properties for future work. More broadly, stellar flaring could potentially affect the thickness and integrated reflectance of the haze present in the water--world exoplanet atmosphere, if no mechanism for production is present. This increased atmospheric loss will affect the climate and ultimately potential habitability of water--world exoplanets to a great extent. Our experiments show that UV irradiation does affect the spectra of both haze samples, providing critical insight into the impacts of higher--energy radiation on close--orbiting exoplanet hazes. Future work will determine if the amplitude of these changes will be observable when they are included in resulting exoplanetary transmission spectra.

\vspace{0.5cm}

The authors gratefully acknowledge M.~S. Marley for supporting this project and to Sukrit Ranjan and Alfred McEwen for providing insightful comments. We also acknowledge the production of the exoplanet haze samples by the Johns Hopkins University PHAZER lab, supported by NASA under the XRP program (Grant 80NSSC20K0271), and support by NASA for this study under the SURP program (Grant 2023--048) between the Jet Propulsion Laboratory and the University of Arizona. This work was supported also by a University of Arizona Postdoctoral Research Development Grant. S.E.M. is supported by NASA through the NASA Hubble Fellowship grant HST-HF2-51563 awarded by the Space Telescope Science Institute, which is operated by the Association of Universities for Research in Astronomy, Inc., for NASA, under contract NAS5-26555. We also acknowledge NASA grants 80NSSC23K0327, NNX12AL47G, NNX15AJ22G and NNX07AI520, and NSF grants 1531243 and EAR--0841669 for funding of the K-ALFAA facility for the optical reflected light mosaic images.

\newpage
\bibliography{bibliography1}

\end{document}